\documentclass[12pt]{article}
\usepackage{amssymb} 
\usepackage{graphicx}
\usepackage{amsmath}

\def\e{{\rm e}}

\def\d{\partial}

\newcommand{\be}{\begin{equation}}
\newcommand{\ee}{\end{equation}}
\newcommand{\bea}{\begin{eqnarray}}
\newcommand{\eea}{\end{eqnarray}}
\newcommand{\bg}{\begin{gather}}
\newcommand{\eg}{\end{gather}}
\newcommand{\bseq}{\begin{subequations}}
\newcommand{\eseq}{\end{subequations}}
\newcommand{\tg}{\mathop{\rm tg}\nolimits}
\newcommand{\arctg}{\mathop{\rm arctg}\nolimits}
\renewcommand{\tanh}{\mathop{\rm th}\nolimits}
\newcommand{\ch}{\mathop{\rm ch}\nolimits}
\newcommand{\sh}{\mathop{\rm sh}\nolimits}
\renewcommand{\ln}{\mathop{\rm ln}\nolimits}

\renewcommand{\Im}{\mathop{\rm Im}\nolimits}
\renewcommand{\Re}{\mathop{\rm Re}\nolimits}

\newcommand{\bra}[1]{\langle #1 |}
\newcommand{\ket}[1]{| #1 \rangle}

\begin{document}
~\bigskip
\begin{center}
  {\LARGE Induced tunneling in QFT: soliton creation
  in collisions of highly energetic particles.} \\
\bigskip
{\large
D.G.~Levkov$^{a,b}$\footnote{Email: \texttt{levkov@ms2.inr.ac.ru}},
S.M.~Sibiryakov$^a$\footnote{Email: \texttt{sibir@ms2.inr.ac.ru}}}\\
\medskip
  $^a${\small Institute for Nuclear Research of the Russian Academy 
    of Sciences,\\
     60th October Anniversary Prospect 7a, Moscow, 117312, Russia}\\
\medskip
$^b${\small Moscow State University, Department of Physics,\\
    Vorobjevy Gory, Moscow, 119899, Russia}
\end{center}

\begin{abstract} 
We consider tunneling
transitions between states separated by an energy barrier 
in a simple field theoretical model. We analyse the case of
soliton creation induced by collisions of a few highly energetic
particles. We present semiclassical, but otherwise first principle,
study of this process
at all energies of colliding particles. We find that 
direct tunneling to the final state
occurs at energies below the critical value $E_c$,
which is slightly higher than the barrier height. Tunneling
probability grows with energy in this regime. 
Above the critical energy, the tunneling mechanism is different.
The transition
proceeds 
through creation of a state close to the top of the potential barrier 
(sphaleron) and its subsequent decay. 
At certain limiting energy $E_l$ tunneling probability
ceases to grow. At higher energies the dominant mechanism of
transition becomes the release of energy excess $E-E_l$ by the
emission of a few
particles and then tunneling at effectively lower energy $E=E_l$ 
via the limiting
semiclassical configuration. The latter belongs to a class
of ``real--time instantons'', semiclassical solutions saturating the
inclusive probability of tunneling from initial states with given number
of particles. We conclude that the process of 
collision--induced tunneling is 
exponentially suppressed at all energies.

\end{abstract}

\section{Introduction}
\label{sec:1} 
Tunneling processes are inherent in many field theoretical models, 
well--known 
examples being false vacuum decay in scalar theories \cite{Coleman:1977py} 
and 
topology--changing transitions in gauge theories
\cite{Belavin:1975fg}. 
To some extent tunneling in field theory is similar to that in
quantum mechanics.
A relevant energy scale for tunneling 
is set by the minimum height of the energy barrier separating 
initial and final states. It is equal to the energy of a critical 
bubble \cite{Coleman:1977py} and sphaleron~\cite{Klinkhamer:1984di} in
the two examples above.  
Classical transition is energetically forbidden at energies below the
barrier height.
In
the weak coupling regime semiclassical technique is applicable for
the description of tunneling, and the rate of transition is exponentially
suppressed. 
On the other hand, one might expect the suppression to vanish once
large enough amount of energy, 
lifting the system above the potential barrier, is injected.
This expectation turns out to be correct in some situations, such as 
transitions at finite temperature (see Ref.~\cite{Rubakov:1996vz} and
references therein), at non-zero fermion density 
\cite{Rubakov:1986nk}, 
or in the presence of heavy particles in the initial 
state~\cite{Rubakov:1985ix,Selivanov:1985vt}.
There is, however, a situation which seems to provide a notable exception
from this rule. This is the case of tunneling induced by collision 
of two highly 
energetic particles. 

A prospect of observation of non--perturbative transitions in particle
collisions
\cite{Ringwald:1990ee,Espinosa:1990qn} provoked extensive discussion in 
literature 
(see \cite{Mattis:1992bj,Tinyakov:1993dr,Rubakov:1996vz} for reviews). 
Semiclassical results \cite{Rebbi:1996zx,Kuznetsov:1997az,Bezrukov:2003er} 
show, however, that this process of induced tunneling remains
exponentially suppressed even when  
the energy of colliding particles exceeds considerably the height of
the potential barrier.  
Furthermore, investigations of 
toy models~\cite{Voloshin:1993dk,Rubakov:1994hz} and unitarity 
arguments~\cite{Zakharov:1991rp,Veneziano:1992rp,Maggiore:1991vi} 
indicate that the process of induced
tunneling should remain exponentially suppressed 
even as the energy of collision tends to infinity. 
Nevertheless, no direct calculation 
of the suppression exponent at very large energies of colliding
particles has been carried out so far in 
any model, not to speak about the limit of infinite energy.

In this paper we consider a simple field theoretical model allowing
for the semiclassical calculation of the
suppression exponent of induced tunneling at {\em all} energies
of colliding particles. 
The model describes free scalar field $\phi(t,{\rm x})$
living in (1+1) dimensions  
on a half--line ${\rm x}>0$, with interactions localized at the
boundary point ${\rm x}=0$. The action of the 
model is
\begin{equation}
\label{eq:S}
S = \frac12\int dt\int\limits_0^\infty d{\rm x}
\left[(\partial_\mu\phi)^2-m^2\phi^2\right] -
\frac{\mu}{g^2}\int dt\left[1-\cos(g\phi(t,0))\right].
\end{equation}
The second term represents boundary interaction with 
the characteristic energy scale $\mu$.
The bulk mass $m$ is introduced as an infrared regulator, and it is
assumed to 
be small compared to the scale $\mu$. 
In the main body of the paper we take the limit $m\to0$, as the small
mass turns out to be irrelevant for the semiclassical evaluation of the 
suppression exponent. The only role it plays 
is to fix
the free particle basis
for the asymptotic states.

The massless version of the model (\ref{eq:S}) with different infrared
regularizations has been considered in several
contexts. In condensed matter physics it has been used to describe
transport in quantum wires \cite{Kane} and Josephson chains with 
defects \cite{Fazio}, the role of the regulator in those cases being played
by the length of wire or chain. In
\cite{Callan:1993mw}, a special value
of the coupling constant $g=\sqrt{2\pi}$ was considered and the model
was shown to be 
exactly solvable in that case. 
Finally, in \cite{Fendley:1994rh} the model (\ref{eq:S})
with $m=0$, obtained as a limit of a model with sine-Gordon
interaction in the bulk, was argued to be integrable 
at any value~of~$g$. 

The model (\ref{eq:S}) possesses a number of static solutions
determined by the value of the field at ${\rm x}=0$. Up to
corrections of order $m/\mu$, these solutions have the following form: 
the value of the field at the boundary is ``pinned down'' to one of
the minima of the boundary
potential, $\phi_0^{(n)}=2\pi n/g$, $n=1,2,\dots$, 
while in the bulk the field slowly falls off due to the
presence of the small mass,
\begin{equation}
\nonumber
\phi_{\mathrm{sol}}^{(n)}({\rm x}) = \phi_0^{(n)}\mathrm{e}^{-m\mathrm{x}}\;.
\end{equation}
These solutions are localized near the boundary, ${\rm x}=0$, so it is
natural to call them ``boundary solitons''. The 
masses of the solitons,
\begin{equation}
\label{eq:Mn}
M^{(n)} =  \frac{2\pi^2n^2}{g^2}m\;,
\end{equation}
are relatively small, as they are proportional 
to the bulk mass $m$.

In this paper we study the process of
creation of the first soliton ($\phi_0^{(1)}=2\pi/g$)
in a collision of a particle (or a bunch of particles) 
with the boundary\footnote{If the bulk mass $m$ is equal to zero (and
the model is somehow regularized in infrared) the spatially
homogeneous configurations $\phi^{(n)}({\rm x})=\phi_0^{(n)}$ are
classical vacua of the model. Our results then apply to
vacuum--to--vacuum transitions induced by collisions of highly
energetic particle(s) with the boundary. All formulae in the main text
remain valid. Yet another situation where our analysis applies
literally, is induced false vacuum decay in a model with $m=0$, and
degeneracy between the classical vacua $\phi^{(n)}({\rm
x})=\phi_0^{(n)}$ slightly lifted.}; 
the 
energy of the particle(s) is assumed  
to be much larger than the soliton 
mass. To start with, let us consider the classical analog of this
process, which is the creation of the soliton in the collision 
of a  classical wave packet with the boundary. It is easy to see that
the classical process is 
possible only if the energy of the wave 
packet exceeds some threshold energy $E_S$, which is much higher than
the soliton mass (\ref{eq:Mn}). Indeed, the boundary value
$\phi(t,0)$ of  
the relevant classical solution changes from $0$ to $2\pi/g$ during
the process. The classical solution passes the maximum of
the boundary potential $\phi_0^{(S)}=\pi/g$  
at some moment of time, 
and thus the total energy of this solution is larger than the
boundary energy at the maximum,  
\begin{equation}
\label{eq:Es}
E_S = \frac{2\mu}{g^2}\;.
\end{equation}
One concludes that any state
containing boundary soliton is separated from the vacuum by a
potential barrier. 
It is straightforward to
find static configuration ``sitting'' on top of the barrier, which 
we call ``sphaleron''
following Ref.~\cite{Klinkhamer:1984di}. 
This is an unstable static solution of the classical field 
equations representing the saddle point of the static energy functional. 
In our model it has the same exponential form
as the soliton but with the boundary value on top of the boundary
potential,     
\begin{equation}
\nonumber
\phi_S = \frac{\pi}{g} \mathrm{e}^{-m\mathrm{x}}\;.
\end{equation}
The energy of the sphaleron is given by Eq.~\eqref{eq:Es}, again up to
corrections of order $m/\mu$. 

We see that the soliton production in collisions of particle(s) with
the boundary 
is classically forbidden and hence exponentially suppressed at
energies smaller than the sphaleron energy.  
The question is what happens when energy grows.
In this paper we study this question by applying semiclassical methods. The
semiclassical approximation is justified by the following
observation.
After the rescaling of the field, $\phi\to\phi/g$, the coupling constant
$g$ enters the action only through the overall multiplicative 
factor $1/g^2$. Therefore, $g^2$ 
plays the role of the Planck constant $\hbar$, and the weak coupling
limit corresponds to the semiclassical  
situation. This is the case we consider in this paper.

The techniques we use are explained in more detail in subsequent
sections. Here we briefly summarize our results.
We calculate semiclassically the probability 
${\cal P}$ of the soliton production in collision of a
particle 
with the boundary. 
In the leading semiclassical approximation it has the exponential form,
\begin{equation}
\nonumber
{\cal P} \propto \mathrm{e}^{-F/g^2},
\end{equation}
and we concentrate on the calculation of the suppression exponent $F$. 
\begin{figure}[htb]
\begin{center}
\includegraphics[width=0.8\textwidth]{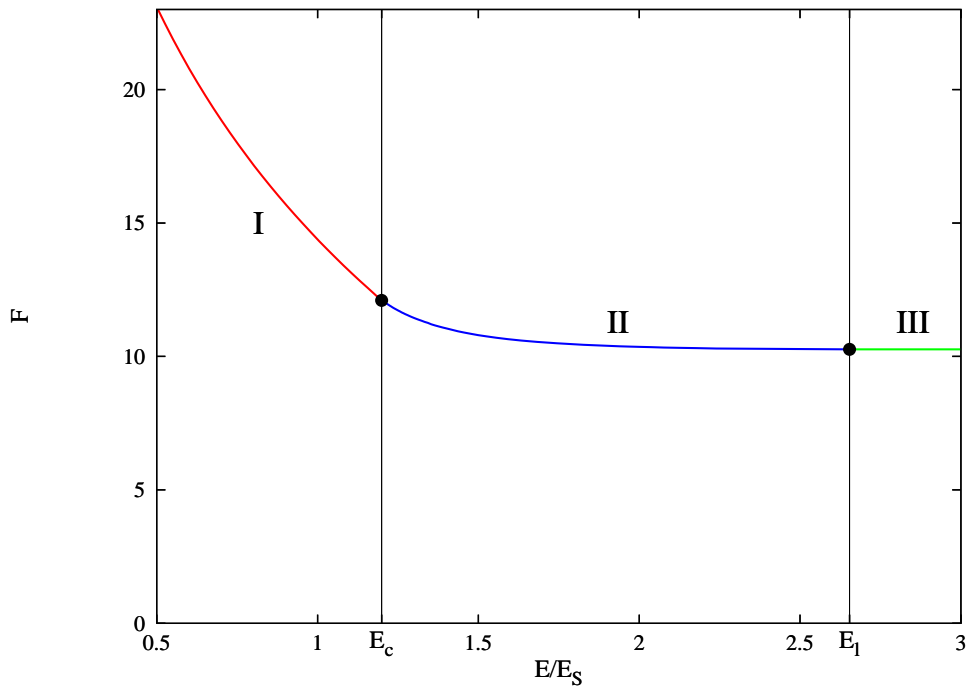}
\end{center}
\caption{Dependence of the suppression exponent $F$ on the collision energy
$E$, the latter measured in units of  
the sphaleron energy $E_S = 2\mu/g^2$. The tunneling process 
is driven by three physically different mechanisms in the parts I, II,
III of
the graph.} 
\label{fig:1}
\end{figure}
Our results for the dependence of the suppression exponent
on the energy of incoming particle are shown in Fig.~\ref{fig:1}.
There are three energy regions 
corresponding to three physically different mechanisms of the process.
If the collision
energy is lower than the critical value $E_c\approx 1.2E_S$ 
(region I of the figure), 
tunneling occurs in conventional way with the
relevant semiclassical configurations ending up directly in the soliton
sector. This regime can be called ``direct
tunneling''. 
We study this region of energies in Sec.~\ref{sec:direct}, and
find the analytic result for the suppression of such transitions:
\begin{equation}
\label{eq:E1}
F(E) = 4\pi\ln\left[\frac{\pi E_S}{E}\right],\;\;\; E < E_c\;. 
\end{equation}
Formula (\ref{eq:E1}), if continued to the energies higher than
the critical energy $E_c$, would show that the 
transitions 
become unsuppressed at energy $\pi E_S$. 
However, this formula is incorrect at $E > E_c$,
as the solutions describing direct tunneling cease to
exist at energies higher than the critical one. 

Semiclassical configurations with energies above $E_c$ are studied in
Sec.~\ref{sec:jumps} using a combination of analytic and numerical methods. 
We find that their properties are different from the ones
below the critical energy, and 
the tunneling mechanism they describe is entirely different. 
Instead of tunneling directly to the other side of the energy barrier, the
system jumps on its top, thus creating  
the sphaleron 
which then decays producing the soliton
in the final state. The second stage of the process, decay of the
sphaleron into the soliton, proceeds with the probability of order
one. 
Still, the transitions remain
exponentially suppressed due to considerable rearrangement the system
has to undergo during the first stage of the process, i.e. formation
of the sphaleron. The tunneling mechanism outlined here has been
observed recently in quantum mechanics \cite{Bezrukov:2003tg} and
gauge theory \cite{Bezrukov:2003er}.
Transitions of this type 
correspond to the region II of Fig.~\ref{fig:1}, 
they extend up to the energy $E_l\approx 2.7\;E_S$. 

The solution describing the transition with the 
energy $E_l$ belongs to a novel class
of semiclassical solutions which we call ``real--time instantons''.
The latter saturate the
inclusive probability of tunneling from initial states with given number
of particles. We consider the real--time instantons in 
Sec.~\ref{sec:real-time}.

The energy $E_l$ is the best energy for the transition.
Further increase of energy above $E_l$
(region III of Fig.~\ref{fig:1}) does not lead to any gain in
the transition probability,  
and
the suppression exponent stays constant all the way up to
$E=\infty$. 
We show in Sec.~\ref{sec:emission} that 
the system emits the 
energy excess $(E-E_l)$ in the form of 
a few highly energetic particles, and then
undergoes the transition  
of the second type (on top of the energy barrier) 
with the energy $E_l$. Emission of one or several
particles does not change the suppression exponent, affecting 
only the pre-exponential factor. In this way we obtain that 
$F(E>E_l)= 10.27$.  
Our results confirm the conjecture
\cite{Maggiore:1991kh}
that the suppression exponent of collision--induced tunneling is frozen at
high energies. The tunneling mechanism at $E>E_l$, i.e.
emission of energy excess in the form of a few particles and tunneling
via the most favourable semiclassical configuration, coincides with
that conjectured in Ref.~\cite{Voloshin:1993dk}.

Section \ref{sec:discussion} contains concluding remarks.

\section{$T/\theta$--problem}
\label{sec:2}
\subsection{General formalism}
\label{sec:general}

A difficulty one encounters in the semiclassical description of  
collision--induced 
tunneling is that the initial state of the process is not
semiclassical. To this end, 
the method of calculation should be appropriately adjusted. In this work we 
adopt the method proposed in~\cite{Rubakov:1992ec}. Namely, let us
consider the inclusive probability of tunneling 
from multiparticle states with energy $E$ and number of particles $N$:
\begin{equation}
\label{eq:PEN}
{\cal P}(E,N) = \sum\limits_{i,f} \left|\langle f| \hat{\cal S}
\hat{P}_E\hat{P}_N|i\rangle\right|^2\;,
\end{equation}
where $\hat{\cal S}$ is the $S$-matrix while $\hat{P}_E$ and $\hat{P}_N$ are 
projectors onto states with given energy and number of particles. The states 
$|i\rangle$ and $|f\rangle$ are perturbative excitations above the vacuum and 
the soliton, respectively. 
The function \eqref{eq:PEN} can be calculated with the use of
semiclassical methods, 
provided that the energy and initial 
number of particles are semiclassically large, $E = \tilde{E}/g^2$, 
$N = \tilde{N}/g^2$. The result has the exponential form,
\begin{equation}
\nonumber
{\cal P}(E,N) \propto \mathrm{e}^{-F(\tilde{E},\tilde{N})/g^2}\;.
\end{equation}
It is clear that the multiparticle probability ${\cal P}(E,N)$ 
provides an upper bound on the tunneling probability 
induced by collisions of fewer particles. Indeed, any initial 
few--particle state can be promoted to a multiparticle
one by adding an appropriate number of ``spectator'' particles which do not
affect the tunneling process. Thus, 
the exponent $F(\tilde{E})$ of the soliton production in the collision
of {\em one} particle with the boundary, is larger than the 
inclusive suppression exponent $F(\tilde{E},\tilde{N})$. 
Furthermore, one conjectures~\cite{Rubakov:1992ec}
that 
the small--particle limit of the multiparticle exponent coincides with 
the exponent of tunneling induced by one or a few particles,
\begin{equation}
\label{eq:FEN}
F(\tilde{E}) = \lim_{\tilde{N}\to0} F(\tilde{E},\tilde{N})\;.
\end{equation}
The relation (\ref{eq:FEN}) has been checked in several orders 
of perturbation theory around the instanton in gauge 
theory~\cite{Tinyakov:1991fn},
and also non-perturbatively in quantum mechanics of two degrees of 
freedom~\cite{Bonini:1999kj,Bezrukov:2003tg}. We use Eq.~\eqref{eq:FEN} 
throughout this paper to obtain $F(\tilde{E})$ from the 
multiparticle exponent $F(\tilde{E},\tilde{N})$. To simplify notations, 
we omit tilde over the rescaled energy and number of particles below.

\begin{figure}[htb]
\centerline{\includegraphics[width=0.8\textwidth]{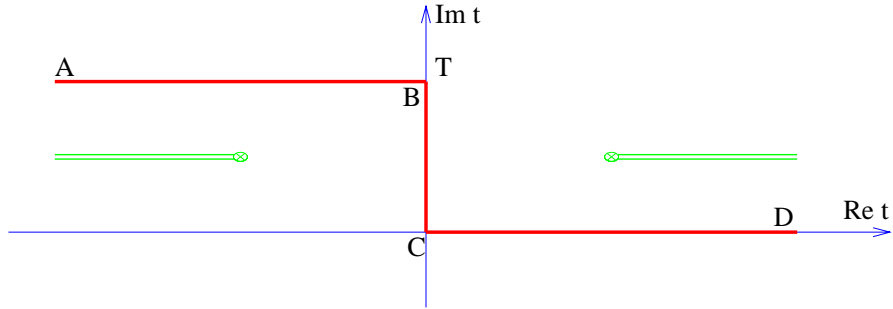}}
\caption{Contour in complex time on which the boundary value 
problem is formulated.}
\label{fig:2}
\end{figure}

Transitions with fixed initial number of particles $N$ are described 
by solutions of the so--called $T/\theta$ boundary value 
problem~\cite{Rubakov:1992ec}. The latter 
is formulated on the contour $ABCD$ in complex 
time plane shown in Fig.~\ref{fig:2}. 
Namely, the configurations describing tunneling should satisfy the
classical field equations
in the internal points of the contour,
\begin{subequations}
\label{eq:eq}
\begin{eqnarray}
\label{eq:eqa}
(\partial^2_t-\partial^2_{\rm x} + m^2) \phi = 0,\;\;{\rm x} > 0\;,\\
\label{eq:eqaa}
\partial_{\rm x} \phi = \mu\sin \phi,\;\;{{\rm x}=0}\;.
\end{eqnarray}
The Euclidean part $BC$ of the contour may be interpreted as representing 
the evolution of the field under the barrier, ``duration'' $T$ of this
evolution is a parameter of solution. Field 
equations~\eqref{eq:eqa},~\eqref{eq:eqaa} are supplemented by initial and 
final boundary conditions in the parts $A$ and $D$ of the contour, 
respectively. 
Namely, the field $\phi(t,{\rm x})$ should be 
real as $t\to +\infty$; it describes the final state of the
semiclassical 
evolution,
\begin{equation}
\label{eq:eqb}
\mathrm{Im}\; \phi \to 0\;\; \mathrm{as}\;\; t\to+\infty\;.
\end{equation}
The field in the initial state is linear about the vacuum $\phi=0$,
\begin{equation}
\label{philinear}
\phi(t'+iT,{\rm x})\Bigg|_{t'\to-\infty} = 
\frac{1}{(2\pi)^{1/2}}\int \frac{dk}{\sqrt{2\omega_k}}\left(
f_k \mathrm{e}^{-i\omega_k t'+ik{\rm x}} + 
g_k^*\mathrm{e}^{i\omega_k t'-ik{\rm x}}\right)\;, 
\end{equation}
and the boundary conditions in the part $A$ 
of the contour relate positive and 
negative frequency components of the solution,
\begin{equation}
\label{eq:eqc}
f_k = \mathrm{e}^{-\theta} g_k\;.
\end{equation}
\end{subequations}
The boundary condition~\eqref{eq:eqc} can be understood as follows. 
In the limit 
$\theta\to +\infty$ it coincides with the Feynman boundary condition and thus 
corresponds to the initial state with semiclassically small number of 
particles. Finite $\theta$ corresponds to 
the initial state with non--zero 
$N$, from which tunneling occurs in the least suppressed way. 

Given the values of $T$ and $\theta$, one finds a complex 
solution $\phi(t,{\rm x}; T,\theta)$ satisfying 
equations~\eqref{eq:eqa} ---~\eqref{eq:eqc}.
The energy and initial 
number of particles for this solution are given by the 
familiar formulae,
\bseq
\label{EN}
\begin{align}
\label{EN1}
&E = \int dk\; \omega_k f_k g_k^*\;,\\
\label{EN2}
&N = \int dk\; f_kg_k^*\;.
\end{align}
\eseq
Alternatively, they can be determined by differentiating the action
functional evaluated on the solution, with respect to the parameters $T$
and $\theta$, 
\begin{equation}
\label{ENLegendre}
E = \frac{\partial}{\partial T} \;\mathrm{Im} S(T,\theta)\;,\;\;\; 
N = 2\frac{\partial}{\partial \theta} \;\mathrm{Im} S(T,\theta)\;,
\end{equation}
where 
\be
\label{action1}
S=\frac{1}{2}\int dt\int_0^{\infty}d{\rm x}~
[-\phi\partial_0^2\phi-(\partial_{\rm x}\phi)^2-m^2\phi^2]-
\mu\int dt~(1-\cos{\phi(t,0)})
\ee
is the action of the 
model\footnote{Hereafter we use the rescaled 
action which does not contain the coupling constant $g$.},
integrated by parts and
calculated along the contour 
$ABCD$.
The suppression exponent of the process is given by the Legendre transform
of the action functional,
\begin{equation}
\label{F}
F(E,N) = 2\Im{S} - N\theta - 2ET\;.
\end{equation}
Below we refer to the problem~\eqref{eq:eq} as ``$T/\theta$-problem'', 
and use the term ``$\theta$--instanton'' for the 
relevant semiclassical solution.

Several remarks are in order. First, the boundary value 
problem~\eqref{eq:eq} does not guarantee that its solutions interpolate 
between states with and without the soliton. In the next subsection we 
describe additional 
requirements that should be imposed to ensure that the solutions are 
relevant
for tunneling. 

Second, one observes that the solution $\phi(t,{\rm x})$ can be analytically 
continued to the complex time plane, and in this way the contour $ABCD$ may 
be deformed without affecting the 
integral~\eqref{F} for the suppression exponent.
The only thing one should worry about while deforming the contour 
is to avoid the 
singularities of the solution, which are shown schematically 
by the double lines in Fig.~\ref{fig:2}.
Below it will be 
convenient not to be attached to a contour of any particular
form. Instead, we look for
the solution $\phi(t,{\rm x})$ 
satisfying Eqs.~\eqref{eq:eqa},~\eqref{eq:eqaa} in the entire complex time 
plane, with boundary conditions~\eqref{eq:eqb} and~\eqref{eq:eqc} 
imposed in the asymptotic regions $D$ and $A$ of the complex plane.
When using this approach, one should make sure, however, that the asymptotic 
regions
$A$ and $D$ can be connected by a contour avoiding the singularities of
solution. 

Finally, let us discuss two particular limits of the boundary 
value problem~\eqref{eq:eq}. The first one corresponds to 
$\theta=0$. Solutions obtained 
in this limit are periodic in Euclidean time, 
they are called ``periodic instantons'' \cite{Khlebnikov:1991th}. 
These solutions determine
the inclusive probability of tunneling from states with given energy
and arbitrary number of particles, 
\[
{\cal P}_{p}(E) = \sum\limits_{i,f}\left|\langle f
|\hat{\cal S}\hat{P}_E|i\rangle\right|^2\;.
\]
Periodic instantons have
been extensively studied in 
literature~\cite{Bonini:2000mb},
in particular, they can be used to determine 
the rate of transitions at finite temperature. 

Below we will encounter solutions of the $T/\theta$-problem 
obtained in the limit
$T=0$. To the best of our knowledge, such solutions 
have never been considered
before. We call them ``real-time instantons'', as the contour $ABCD$
does not contain a Euclidean part in this case. 
These solutions depend on a single auxiliary parameter $\theta$,
i.e. a single free physical parameter of the incoming state, which we
choose to be the number of particles $N$. The energy is thus a
function of $N$ for this set of solutions.
The value of
the suppression exponent $F_{rt}(N)$ evaluated on the 
real-time instantons is 
non-zero exclusively
due to the fact that these field configurations are complex-valued. 
At first sight, the latter property may seem to contradict the condition
(\ref{eq:eqb}) requiring reality of the solution in the region $D$ of the
real axis. However, this is not the case: the condition (\ref{eq:eqb}) is
imposed only in the asymptotic future, and it does not prevent
the solution from being complex--valued at finite times. 
Let us discuss
this point in more detail. Assume that at large (but finite) time
the solution is linearized about some static configuration,
\[
\phi(t,{\rm x})=\phi_0({\rm x})+\delta\phi_1(t,{\rm x})+
i\delta\phi_2(t,{\rm x})\;,
\]
where both $\delta\phi_1$ and $\delta\phi_2$ are real.
The condition (\ref{eq:eqb}) implies that $\delta\phi_2\to 0$ at 
$t \to +\infty$. This amounts to requiring that $\delta\phi_2$
describes the evolution along the negative mode of the static solution
$\phi_0({\rm x})$, so that 
$\delta\phi_2 \propto \e^{-{\rm const}\cdot t}$. 
Thus, the configuration $\phi_0({\rm x})$ must have
negative modes, in other words, it must be unstable. The only
candidate for $\phi_0({\rm x})$ in our case is the sphaleron. We
arrive at the important conclusion that the real--time instantons describe
formation of the sphaleron at $t\to +\infty$. It is worth noting that 
this observation refers, in fact, 
to any $\theta$-instanton which is not real
at real time, cf. Ref.~\cite{Bezrukov:2003tg}.

Let us demonstrate that the real-time instantons determine
the suppression exponent for the inclusive probability of tunneling from
states with given number of incoming particles and {\em arbitrary}
energy 
\be
\label{realP}
{\cal P}(N) = \sum\limits_{i,f}\left|\langle f
|\hat{\cal S}\hat{P}_N|i\rangle\right|^2\;.
\ee
From Eq.~(\ref{eq:PEN}) we obtain,
\[
{\cal P}(N)=\int dE~{\cal P}(E,N)\propto\int dE~\e^{-F(E,N)/g^2}\;.
\]
Let us evaluate the last integral in the saddle point approximation. The
saddle point is determined by condition
\[
\frac{\d F}{\d E}=0\;.
\]
On the other hand, formulae (\ref{ENLegendre}), (\ref{F}) yield
\[
\frac{\d F}{\d E}=-2T\;.
\]
Thus, the integral is saturated by the solution with $T=0$, i.e. real--time
instanton:
\be
\label{realPP}
{\cal P}(N)\propto\e^{-F_{rt}(N)/g^2}\;.
\ee 
The energy $E^{(rt)}(N)$ evaluated on the real--time instanton
solution according to Eq.~(\ref{EN1}), is the best energy for
tunneling at given $N$. We discuss this point further in
Sec.~\ref{sec:emission}.   

Formulae (\ref{realP}), (\ref{realPP}) show that 
$F_{rt}(N)$ sets the lower bound on the suppression exponent $F(E)$
for the one--particle 
collision--induced tunneling at all energies, extending 
up to infinity, and
the smallest value of $F(E)$ coincides with the limit of $F_{rt}(N)$ as
$N\to 0$.

\subsection{Reformulation of the problem}
\label{sec:reformulation}

In this subsection we adapt the $T/\theta$-problem~\eqref{eq:eq} for the 
specifics of the model~\eqref{eq:S}. The key observation here is that the 
characteristic frequency scale 
of solutions under consideration is of the order of the 
boundary scale $\mu$, which is much greater than $m$. 
Thus, the mass term in the bulk 
equation~\eqref{eq:eqa} can be neglected, and the general solution 
in the bulk has the form
\be
\label{decomp}
\phi(t,{\rm x})=\phi_i(t+{\rm x})+\phi_f(t-{\rm x}),
\ee 
where $\phi_i$ and $\phi_f$ are the incoming and outgoing waves, 
respectively. They are related by the boundary condition~\eqref{eq:eqaa}:
\be
\label{boundarynew}
\phi_i'(z)-\phi_f'(z)=\mu\sin{(\phi_i(z)+\phi_f(z))},
\ee
where we have promoted $t$ to complex variable $z$.
It is natural to consider $\phi_i$ and $\phi_f$ as 
analytic functions of $z$, and reformulate the 
rest of the problem~\eqref{eq:eq}, i.e. 
conditions~\eqref{eq:eqb} and~\eqref{eq:eqc}, in the complex $z$--plane. 

Let us consider the asymptotic future, 
$t\to +\infty$ (region $D$ in Fig~\ref{fig:2}). 
In this limit the field $\phi(t,{\rm x})$ 
is represented by the outgoing wave 
$\phi_f(t-{\rm x})$ whose argument $z=t-{\rm x}$ 
runs all the way along the real axis as 
${\rm x}$ changes from $0$ to $+\infty$. So, 
the condition~\eqref{eq:eqb} can be written in the following way,
\begin{equation}
\label{eqf}
\mathrm{Im}\; \phi_f (z)= 0,\;\;\mathrm{when}\;\;z\in \mathbb{R}\;. 
\end{equation}
On the other hand, it is the incoming wave  
$\phi_i(t+{\rm x})$ which survives in the asymptotic 
past $t\to -\infty+iT$ (region $A$ of Fig.~\ref{fig:2}), where 
the argument of 
the function $\phi_i(z)$ runs along the line $\mathrm{Im}\;z = T$ when
${\rm x} \in [0,+\infty)$.   
So,  
the condition~\eqref{eq:eqc} is reformulated 
in terms of the function $\phi_i$. Namely, 
one performs the Fourier expansion of $\phi_i$ along the line 
$\mathrm{Im}\;z = T$,
\begin{equation}
\label{Four}
\phi_i(z) = \int dk\; \phi_i(k) \mathrm{e}^{ik(z-iT)} = 
\int\limits_{k>0} dk\; \left\{\phi_i(k)\mathrm{e}^{ik(z-iT)} +
\phi_i(-k)\mathrm{e}^{-ik(z-iT)}\right\}.
\end{equation}
Taking into account that $z=t'+iT+{\rm x}$~~for the initial wave, one
compares Eq.~(\ref{Four}) to Eq.~\eqref{philinear} and finds that 
the positive and negative frequency components $f_{-k}$ and $g_{-k}^*$
of solution are proportional to $\phi_i(-k)$ and $\phi_i(k)$, $k>0$, 
respectively. Equation \eqref{eq:eqc} takes the form
\[
{\phi_i}(-k)=\e^{-\theta}[\phi_i(k)]^*~,~~~k>0.
\]
Given this condition, the function $\phi_i$ can be represented as
\be
\label{phichi}
\phi_i(z)=\chi(z-iT)+\e^{-\theta}[\chi(z^*+iT)]^*,
\ee
where the function  
\be
\label{chi}
\chi(z)=\int_{0}^{\infty}dk~{\phi_i}(k)\e^{ikz}
\ee
is regular in the upper half plane of its complex 
argument. Equation~\eqref{phichi} provides an
alternative 
formulation of the 
$\theta$--boundary condition (\ref{eq:eqc}).
Note that if the number of incoming particles is finite, the ingoing
wave packet is localized in space. This implies
\be
\label{phiias}
\phi_i(z)\to 0~,~~~~z\to\pm\infty+iT\;.
\ee 
[The latter condition is violated in the case of periodic instantons, see 
Sec.~\ref{sec:periodic}.]
To summarize, the $T/\theta$--problem is 
now given by equations~\eqref{boundarynew}, \eqref{eqf} and \eqref{phichi}
formulated in the complex $z$--plane. 

To make sure that a solution of the above problem is relevant for
tunneling, one should check that the value of the field 
at the boundary ${\rm x}=0$ has correct asymptotics in 
the beginning and the end of the process.
Namely, in the case of direct tunneling with the soliton in the final
state,  
$\phi(t,0)$ must change from $0$ to $2\pi$ as $t$ changes from
$-\infty$ to $+\infty$. In Sec.~\ref{sec:jumps} we
encounter configurations describing creation of the sphaleron at $t\to
+\infty$; 
in that case $\phi(t,0)$ changes from $0$ to $\pi$. 
Thus, one obtains the following conditions:
\bseq
\label{sol*}
\begin{align}
\label{sol1}
&\phi_i(z)+\phi_f(z)\to0~~~\;\;\mathrm{as}\;\;z\to -\infty+iT,\\
\label{sol2}
&\phi_i(z)+\phi_f(z)\to2\pi~~ {\rm or} ~~\pi~~\;\;\mathrm{as}\;\;
z\to +\infty.
\end{align} 
\eseq

Finally, let us analyze the issue of existence of a time contour
connecting the asymptotic regions $A$ and $D$ in Fig.~\ref{fig:2}.
To this end,
note that any singularity $z_i^s$ of the function $\phi_i(z)$ produces a 
whole half--line of singularities $t^s = z_i^s-{\rm x}$ in the complex
time plane, 
which starts from the point $z_i^s$ and extends to the left parallel 
to the real time axis, see Fig.~\ref{fig:2}. One requires that all
these half--lines of singularities of the function $\phi_i(t+{\rm x})$  
in the strip $\Im t\in [0,T]$ 
in complex 
time plane, should be located to the 
left of the relevant contour. Analogously, one observes that 
all the singularities of the function $\phi_f(t-{\rm x})$ in this strip 
should be located to the 
right of the relevant contour. It is clear that one will always be able to 
find some 
contour connecting the regions $A$ and $D$ which
leaves the singularities of $\phi_i(t+{\rm x})$ and 
$\phi_f(t-{\rm x})$ 
to the 
left and right of it respectively, provided that the singularities of
these two functions
do not coincide.  In terms of the variable $z$ this amounts to requiring that 
the singularities, inside the strip 
$\mathrm{Im}\; z\in[0,T]$, of the initial and final 
wave packets $\phi_i(z)$ and $\phi_f(z)$ are situated at different points.
Note that this condition is non-trivial, as the functions $\phi_i$ and 
$\phi_f$ are related by the differential equation~\eqref{boundarynew}.

Formula~\eqref{action1} for the action can also be rewritten in terms of 
the incoming and outgoing waves:
\be
\label{actionnew}
S=\int_{\cal C} dz \left\{\frac{1}{2}(\phi_i+\phi_f)(\phi_i'-\phi_f')-
\mu(1-\cos(\phi_i+\phi_f))\right\}.
\ee
The form of the contour ${\cal C}$ here is somewhat 
similar to the time contour $ABCD$ in 
Fig.~\ref{fig:2}: it interpolates between  
the asymptotic regions $z\to-\infty+iT$ and
$z\to+\infty$, leaving the singularities of the 
functions $\phi_i$ and $\phi_f$
in the strip $\Im z\in [0,T]$
to the left and right, respectively.

\section{Direct soliton production at low energies}
\label{sec:direct}

\subsection{Periodic instantons}
\label{sec:periodic}
To warm up, let us consider 
periodic instantons which are solutions of 
the problem with $\theta=0$. 
One begins by noticing that Eq.~(\ref{boundarynew}) 
possesses a solution, which would be a conventional instanton in
this model,
\be
\label{instanton}
\phi_i=i\ln\left[\frac{\mu}{2}\left(z+\frac{1}{\mu}\right)\right]~,~~~~
\phi_f=-i\ln\left[\frac{\mu}{2}\left(z-\frac{1}{\mu}\right)\right]\;.
\ee
The corresponding field configuration $\phi$, 
Eq.~(\ref{decomp}), is real along
Euclidean time axis $t=-i\tau$, it changes from $0$ to
$2\pi$ as $\tau$ runs from $-\infty$ to $+\infty$.
So, the properties of the configuration (\ref{instanton}) are  
those one expects from an 
instanton describing vacuum--to--vacuum tunneling.
However, Euclidean action of the instanton (\ref{instanton})
diverges, implying that 
tunneling at zero energy is absent in
our case. 
Obviously, a configuration obtained
from (\ref{instanton}) by the overall change of sign is also a solution, 
which we call anti-instanton.

Exact periodic instanton in our model can be constructed with the
help of an Ansatz 
inspired by the dilute instanton gas approach
\cite{Rubakov:1992ec,Khlebnikov:1991th,Son:1993bz}. Let us take
a periodic chain of alternating instantons and 
anti-instantons\footnote{Constants $3\pi/2$ and $\pi/2$ in 
Eqs.~(\ref{PIfirst}) are added for convenience.}, 
\bseq
\label{PIfirst}
\begin{align}
\label{PIfirst1}
&\phi_i=\frac{3\pi}{2}+\sum_{n=-\infty}^{\infty}\left(
i\ln\left[\frac{\mu}{2}\left(z+x_0-i\frac{T}{2}-i2Tn\right)\right]
-i\ln\left[\frac{\mu}{2}\left(z+x_0+i\frac{T}{2}-i2Tn\right)\right]\right)\;,\\
&\phi_f=\frac{\pi}{2}+\sum_{n=-\infty}^{\infty}\left(
-i\ln\left[\frac{\mu}{2}\left(z-x_0-i\frac{T}{2}-i2Tn\right)\right]
+i\ln\left[\frac{\mu}{2}\left(z-x_0+i\frac{T}{2}-i2Tn\right)\right]\right)\;.
\end{align} 
\eseq
It is straightforward to check that the functions $\phi_i$, $\phi_f$ given
by Eqs.~(\ref{PIfirst}) are real along the lines 
$\Im z=0$ and $\Im z=T$, so the conditions (\ref{eqf}), (\ref{phichi}) are
satisfied with $\theta =0$.  
Note that (anti-)instantons in the chain (\ref{PIfirst}) 
are modified by the introduction of 
a real parameter $x_0$ which is to be determined from the boundary equation of
motion (\ref{boundarynew}).
By making use of the infinite product representation for the hyperbolic sine
(see e.g. Ref.~\cite{Ryzhik}, Eq.~1.431.2), 
summation in (\ref{PIfirst}) can be performed explicitly:
\be
\label{PIsecond}
\phi_i=\frac{3\pi}{2}+i\ln\left(\frac{\sh\left(\frac{\pi(z+x_0)}{2T}-
i\frac{\pi}{4}\right)}{\sh\left(\frac{\pi(z+x_0)}{2T}+
i\frac{\pi}{4}\right)}\right)~,~~~~
\phi_f=\frac{\pi}{2}-i\ln\left(\frac{\sh\left(\frac{\pi(z-x_0)}{2T}-
i\frac{\pi}{4}\right)}{\sh\left(\frac{\pi(z-x_0)}{2T}+
i\frac{\pi}{4}\right)}\right)\;.
\ee
Inserting expressions (\ref{PIsecond}) into Eq.~(\ref{boundarynew}), we
find that the Ansatz satisfies this equation, provided
that $x_0$ satisfies the following relation,
\[
\frac{\mu T}{\pi}\tanh\left(\frac{\pi x_0}{T}\right)=1\;.
\]
Choosing the contour of integration in expression (\ref{actionnew})
as prescribed in Sec.~\ref{sec:reformulation}, one evaluates the 
imaginary part
of the action:
\[
2\Im S=4\pi\ln\frac{\mu T}{\pi}+4\pi\;.
\] 
Now, it is straightforward to obtain the energy and suppression
exponent in the case of periodic instanton from
Eqs.~(\ref{ENLegendre}), (\ref{F}),
\begin{gather}
\label{PIenergy}
E=\frac{2\pi}{T}\;,\\
\label{PIsuppression}
F=4\pi\ln\left(\frac{2\mu}{E}\right)\;.
\end{gather}
In accordance with the above discussion,  
the suppression exponent 
starts from infinity at zero energy. As one could expect,
it vanishes at
the sphaleron energy $E_S=2\mu$.

Let us work out in more detail the structure of the initial and final
states of the processes described by the periodic instantons. In the initial
region ($t=iT+t'$, $t'\to -\infty$) one has
\be
\label{inpacket}
\phi(t',{\rm x})=\pi+\phi_i(iT+t'+{\rm x})=
2\arctg\left[\exp\left(\frac{\pi (t'+{\rm x}+x_0)}{T}\right)\right]\;.
\ee 
This wave packet has the shape of a kink solution of the sine-Gordon
theory with potential $V\propto \cos(2\phi)$. 
At the same time, the outgoing wave ($t\to +\infty$ along the real axis),
\[
\phi(t,{\rm x})=\phi_f(t-{\rm x})=
\pi+2\arctg\left[\exp\left(\frac{\pi (t-{\rm x}-x_0)}{T}\right)\right]\;,
\]
has the form of antikink. Scattering of such 
kink-shaped wave packets from the boundary 
in the framework of massless version of the model (\ref{eq:S})
has been considered in Ref.~\cite{Fendley:1994rh}, and an exact
expression for the amplitude of  kink--to--antikink scattering  
has been obtained there. Our expression (\ref{PIsuppression})
matches the results of \cite{Fendley:1994rh} in the weak coupling
regime.

Finally, let us note that in the strictly massless case, the number of
particles both in the initial and final states of the process
described by the periodic
instanton is infinite, as 
the field has non-zero asymptotics at spatial infinity, 
$\phi(t,{\rm x}\to +\infty)=\pi$.
As shown in Appendix~\ref{app:A0},  
introduction of the bulk mass regularizes this divergence and enforces
the asymptotics $\phi(t,{\rm x}\to +\infty)=0$. 
The corresponding corrections to the suppression exponent 
(\ref{PIsuppression}) are small provided the
energy of the initial state is much higher than the mass $m$.

\subsection{Solutions at arbitrary $\theta$}

We now proceed to solutions with $\theta\neq 0$. Let us
make the following observation: if $\phi_i$ 
is {\em real} on the real axis, the function $\phi_f$ determined
from Eq.~(\ref{boundarynew}) with real initial condition, 
is automatically real on the real axis. 
Thus, all we need is an Ansatz for $\phi_i$,
which satisfies the condition (\ref{phichi}) and is real on the real axis.
One constructs the required Ansatz as an appropriate generalization of
Eq.~(\ref{PIfirst1}), 
\be
\label{phiiAnsatz}
\phi_i=\sum_{n=-\infty}^{+\infty}\e^{-\theta |n|}\left(
i\ln\left[\frac{\mu}{2}(z+x_0-iT_0-i2Tn)\right]
-i\ln\left[\frac{\mu}{2}(z+x_0+iT_0-i2Tn)\right]\right)\;,
\ee
where $x_0$, $T_0$ are real parameters, with $0<T_0<T$. 
It is straightforward to check that the function $\phi_i$
determined by (\ref{phiiAnsatz}) 
can be represented in the form (\ref{phichi}). 

Let us show that (\ref{phiiAnsatz})  
is the most general Ansatz for the function $\phi_i$, once 
the requirement of
its reality on the real axis 
is imposed. 
The function $\phi_f$ must be regular at the 
singularity point of $\phi_i$ situated in the 
strip $0<\Im z<T$. This guarantees that the singularity of
$\phi_i$ in this strip is logarithmic; 
we postpone the proof of this statement till
Sec.~\ref{sec:jumps}.   
Then, reality
on the real axis together with the $\theta$-condition (\ref{phichi}) 
fix all other singularities of $\phi_i$. 
Namely, due to
Eq.~(\ref{phichi}) any singularity of the function $\phi_i$ 
of the form
$$
C\ln(z-z_s)\;,
$$
situated below
the line $\Im z=T$ is accompanied by a 
singularity 
\[
e^{-\theta}C^*\ln(z-2Ti-z_s^*)
\]
placed symmetrically with respect to
the line $\Im z=T$. At the same time, reality on the real axis implies
that all the singularities of $\phi_i$ arise in ``complex
conjugate'' pairs situated at conjugate points. These two conditions
are sufficient to reconstruct the whole chain
(\ref{phiiAnsatz})
from the logarithmic singularity in the strip $0<\Im z<T$. 

The parameter $x_0$ in Eq.~(\ref{phiiAnsatz}) can be
removed by a shift of the variable $z$,  and without
loss of generality we set
$x_0=0$. 
In order to determine the remaining parameter $T_0$ of the Ansatz, 
let us analyze 
equation (\ref{boundarynew}) 
in the vicinity of
the point $z=iT_0$. 
One represents the function $\phi_i$ in the form
\be
\label{Ri}
\phi_i=i\ln\left[\frac{\mu}{2}(z-iT_0)\right]+R_i(z)\;,
\ee
where $R_i(z)$ is regular at $z=iT_0$. 
Power series
expansion in
Eq.~(\ref{boundarynew}) near the point $z=iT_0$ has the form
\be
\notag
\begin{split}
\frac{i}{z-iT_0}&+R_i'(iT_0)-\phi_f'(iT_0)+O(z-iT_0)\\
&=-i\e^{i(R_i(iT_0)+\phi_f(iT_0))}
\left[\frac{1}{z-iT_0}+iR_i'(iT_0)+i\phi_f'(iT_0)\right]+O(z-iT_0)\;.
\end{split}
\ee
Two leading terms of this equation yield
\begin{gather}
\label{RiT_0}
\e^{i(R_i(iT_0)+\phi_f(iT_0))}=-1\;,\\
\label{RiT_0'}
R_i'(iT_0)=0\;.
\end{gather}
These formulae deserve a comment. Considering Eq.~(\ref{boundarynew})
with given $\phi_i$ as
an ordinary differential equation for the function $\phi_f$, 
one might expect 
all Taylor
coefficients of $\phi_f$ to be determined in terms of 
the function $R_i(z)$
and
free integration constant $\phi_f(iT_0)$.
However, the fact that $z=iT_0$ is a singular point of 
Eq.~(\ref{boundarynew}) makes the situation quite different.
The requirement of regularity of
$\phi_f$ at this point fixes the value of
$\phi_f(iT_0)$ according to Eq.~(\ref{RiT_0}), 
while the role of the integration constant is played 
by $\phi_f'(iT_0)$; besides, the constraint (\ref{RiT_0'}) on 
the function $R_i$
appears. 
The latter constraint enables one to determine the parameter $T_0$,
\be
\label{TT0}
T_0=T\alpha(\theta)\;,
\ee
where the function $\alpha(\theta)$ is implicitly defined by 
\be
\label{alphasum}
2\alpha^2\sum_{n=1}^{\infty}\frac{e^{-\theta n}}{n^2-\alpha^2}=1\;.
\ee
This relation can be cast into a convenient integral 
form\footnote{One observes that 
the sum $I(\alpha,\theta)$ in Eq.~(\ref{alphasum}) 
satisfies the differential equation
$$
{\frac{\d^2 I}{\d\theta^2}=
\alpha^2 I+\frac{1}{e^\theta-1}\;.}
$$
By solving this equation 
with appropriate boundary conditions, one obtains (\ref{alpha}).
}
\be
\label{alpha}
2\alpha\int_{0}^{\infty}\frac{\sh(\alpha y)}{e^{y+\theta}-1}dy=1\;.
\ee
The function $\phi_i$ is now completely fixed, and $\phi_f$ can be
obtained by numerical integration of Eq.~(\ref{boundarynew}). We
return to the evaluation of $\phi_f$ at the end of this subsection. 

Let us evaluate the imaginary part of the
action on the tunneling solution. 
Surprisingly, the detailed knowledge of $\phi_f$ is not needed for this
purpose. 
Reality of the solution on the real axis implies that the complex conjugate 
action $S^*$ is given by an integral of the same function as in
Eq.~(\ref{actionnew}), 
but with different contour of integration ${\cal C}^*$, which is complex
conjugate to ${\cal C}$. Thus, 
\be
\label{imaction}
2\Im S=-i(S-S^*)=-i\left(\int_{\cal C}{\cal L}dz-
\int_{{\cal C}^*}{\cal L}dz\right)
=-i\oint_{{\cal C}_o}{\cal L}dz\;.
\ee
In the last equality we deformed the sum of the contours ${\cal C}$
and ${\cal C}^*$ into the contour ${\cal C}_o$ enclosing 
the singularities\footnote{In the course of the deformation, the contour does
not cross the singularities of $\phi_f$, according to the discussion
in Sec.~\ref{sec:reformulation}.}
$z=\pm iT_0$
of the function $\phi_i$. The calculation of the integral
(\ref{imaction}) is now straightforward: 
\[
2\Im S=4\pi \Im\phi_f(iT_0)+4\pi\;.
\]
Using Eq.~(\ref{RiT_0}) one substitutes $R_i(iT_0)$ for
$\phi_f(iT_0)$,  
takes the former from Eq.~(\ref{phiiAnsatz}) and
performs summation. The result is
\be
\label{Stheta}
2\Im S=4\pi\ln(\mu T \alpha)+4\pi-16\pi\int_0^{\infty}
\frac{\sh^2\left(\frac{\alpha y}{2}\right)}{e^{y+\theta}-1}\frac{dy}{y}\;.
\ee 
The energy and number of incoming particles are
determined from expression (\ref{Stheta}) in the standard way, 
see Eqs.~(\ref{ENLegendre}).
We find that the energy is given by
the same formula (\ref{PIenergy}) as in the periodic instanton case,
while the number of incoming particles is
\[
N=4\pi\int_0^{\infty}
\frac{\sh^2\left(
\frac{y\alpha (\theta)}{2}\right)}{\sh^2\left(\frac{y+\theta}{2}\right)}
\frac{dy}{y}\;.
\]

Let us consider in detail the limit $\theta\to +\infty$ which is of
primary interest. One obtains
\begin{align}
\label{alphaN0}
&\alpha=1-e^{-\theta}\;,\\
\label{NN0}
&N=4\pi \theta e^{-\theta}~,~~~~~~\theta\to+\infty\;.
\end{align}
As expected, the limit of large $\theta$ corresponds to
tunneling induced by a few incoming particles. 
The value of the suppression exponent is given by a simple formula in
this limit, 
\be
\label{FN0}
F\big|_{N=0}=4\pi\ln\left[\frac{2\pi\mu}{E}\right]\;.
\ee
This is the result we claimed in Introduction for the region I in
Fig.~\ref{fig:1}. 

At first sight, Eq.~(\ref{FN0}) suggests that the suppression vanishes 
when the 
energy reaches the value 
$2\pi\mu$. In fact, this is not the case: formula
(\ref{FN0}) 
is inapplicable at energies above some critical energy
$E_c<2\pi\mu$. The point is that no solution $\phi_f(z)$ of 
Eq.~(\ref{boundarynew})
with required properties exists at energies $E>E_c$. 
Let us clarify this point.

\begin{figure}[htb]
\begin{center}
\includegraphics[height=6cm,width=0.8\textwidth]{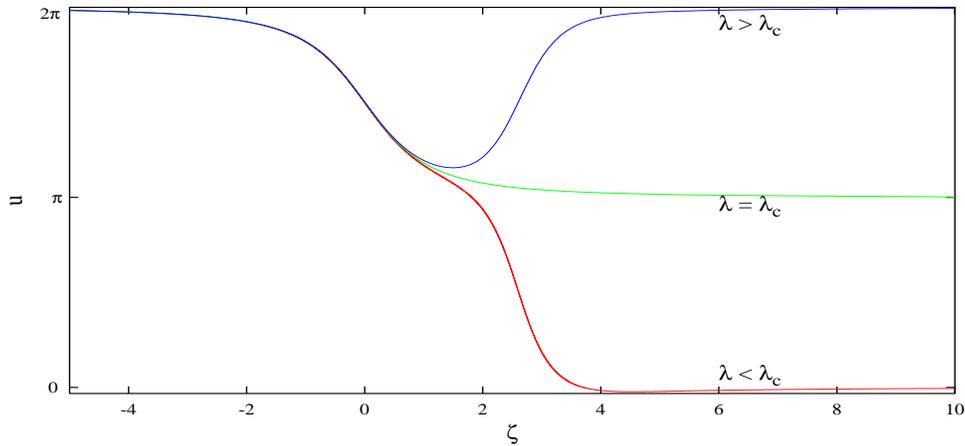}
\end{center}
\caption{Solutions of Eq.~(\ref{ueq}), found numerically 
for $\lambda=2.63$ (upper curve),
$\lambda=\lambda_c=2.62$ (middle curve), $\lambda=2.61$ (lower curve).} 
\label{Fig:u}
\end{figure}

While the analysis can be carried out for arbitrary $\theta$, it is
particularly transparent in the case $\theta=+\infty$. 
According to Eq.~(\ref{alphaN0}), 
the logarithmic singularity of $\phi_i$ approaches the ``incoming
wave line'' $\Im z =T$
in this limit\footnote{\label{foot}This is a generic
feature of $T/\theta$ instantons: Eqs.~(\ref{EN}) show that by
sending the number of incoming particles to zero while keeping their
energy fixed, one obtains configurations 
with infinitely high frequencies and thus
singular on the initial part of the
contour $ABCD$ of Fig.\ref{fig:2}. We overcome this problem  
by considering the functions $\phi_i$ and $\phi_f$ in the whole complex
plane.}, 
and the
Ansatz (\ref{phiiAnsatz}) simplifies:
\[
\label{phii0}
\phi_i=i\ln\frac{z-iT}{z+iT}\;.
\]
It is convenient to consider Eq.~(\ref{boundarynew}) on the real
axis. Introducing $u=\phi_i+\phi_f$ and $\zeta=z/T$,
one writes Eq.~(\ref{boundarynew}) in the following form,
\be
\label{ueq}
\frac{du}{d\zeta}=-\lambda\sin{u}-\frac{4}{\zeta^2+1}\;,
\ee 
where 
$$
\lambda=\mu T\;.
$$ 
In new terms, the requirements (\ref{sol*}), ensuring that the 
solution
is relevant for the soliton production, 
imply the following boundary conditions for
$u$ along the real axis:
\bseq
\label{ubound*}
\begin{align}
\label{ubound1}
&u\to 2\pi~,~~~\zeta\to -\infty\;,\\
\label{ubound2}
&u\to 2\pi~~~{\rm or}~~~\pi,~~~\zeta\to +\infty\;,
\end{align}
\eseq
where the condition (\ref{ubound1}) follows from 
Eq.~(\ref{sol1}) when one takes into account that the asymptotic region
$z\to -\infty+iT$ and the real axis lie on different sides of 
the logarithmic cut of the function $\phi_i$.
The condition (\ref{ubound1}) fixes the solution of
Eq.~(\ref{ueq}) uniquely. Then, the question is whether this
solution satisfies the boundary condition (\ref{ubound2}). 
We have analyzed this issue both analytically
and numerically.
Appendix~\ref{app:A} contains an analytic proof of the existence of the
critical value $\lambda_c$ of the parameter $\lambda$. When $\lambda$
is greater than $\lambda_c$,
the solution of Eqs.~(\ref{ueq}), (\ref{ubound1}) 
has the correct asymptotics (\ref{ubound2}), so the 
configuration indeed describes the production of the soliton. 
On the other hand, when $\lambda$ is below the critical value 
$\lambda_c$,
the asymptotics of the function $u$ at $\zeta\to
+\infty$ change to $0$. These results are confirmed by numerical
analysis, see
Fig.~\ref{Fig:u}.
As $u$ represents the value of the field at
the boundary ${\rm x}=0$ (cf. Eq.~(\ref{decomp})), 
one concludes that the final
state of the process at $\lambda<\lambda_c$ is the trivial vacuum, 
and no soliton production takes
place. 
At the critical value, $\lambda=\lambda_c$, the function $u$ tends to
$\pi$ when $\zeta\to +\infty$, and the
solution describes the production of
the sphaleron in the final state of the semiclassical evolution. 
Numerically we find
\be
\label{lambdac}
\lambda_c=2.62\;,
\ee
and the value of the corresponding energy is $E_c=1.2E_{S}$. 
Tunneling at the critical energy is still exponentially suppressed:
\[
F(N=0,E=E_c)=4\pi\ln{\lambda_c}\;.
\] 
The picture we encounter here has been 
observed recently in quantum mechanics of 
two degrees of freedom
\cite{Bezrukov:2003tg} 
and in gauge theory \cite{Bezrukov:2003er}. 
Results obtained in both cases indicate that
transitions at energies higher than critical proceed in two stages,
creation of the sphaleron and its subsequent quantum decay into the
relevant 
final state. The probability of the latter process is of order one, while
the former is exponentially suppressed. The corresponding semiclassical 
solutions  
contain the sphaleron at $t\to +\infty$. In order to find such solutions
one has to abandon
the requirement of reality of the incoming wave $\phi_i$ on the
real axis and thus the Ansatz (\ref{phiiAnsatz}).

\section{Jumps onto sphaleron}
\label{sec:jumps}

\subsection{Solutions with ``zero'' number of incoming particles}
\label{sec:jumps1}

At energies higher than $E_c$ the Ansatz (\ref{phiiAnsatz}) is no
longer applicable. Still, one can extract some information on the  
properties of
$\phi_i$ from the analysis of Eqs.~(\ref{boundarynew}), (\ref{phichi}). 
First, let us show that the singularity of $\phi_i$ situated  
inside the strip $\Im z\in [0,T]$ is necessarily
logarithmic. We denote the position of the singularity by $iT_0$. 
The function $\phi_f$ is regular at the point $z=iT_0$, so $\phi_f$ and
$\phi_f'$ can be replaced by constants in small vicinity of this
point. Then, integration of Eq.~(\ref{boundarynew}) yields
\be
\label{arct}
\phi_i(z)=-f_0+2\arctg\left[\frac{\sqrt{f_1^2-1}}{f_1}
\tg\left(\frac{\sqrt{f_1^2-1}}{2}(\mu
z-C)\right)-\frac{1}{f_1}\right]~,~~~~z\approx iT_0\;,
\ee
where $f_0=\phi_f(iT_0)$, $f_1=\phi_f'(iT_0)/\mu$, and $C$ is the
integration constant fixed by the condition that $z=iT_0$ is the singular
point of the function $\phi_i$. One observes that
the singularity of the expression (\ref{arct}) is logarithmic.

Thus the function $\phi_i$ has the form (\ref{Ri}) with the function $R_i$
regular in the strip $\Im z\in [0,T]$.
Besides, one observes that
conditions (\ref{RiT_0}), (\ref{RiT_0'}) on $R_i$ are still valid, as their
derivation   
does not make use of any particular
Ansatz for $\phi_i$. Finally, due to Eq.~(\ref{phichi}), 
the presence of the logarithmic singularity of $\phi_i$
at $z=iT_0$ implies the existence of another 
singularity of $\phi_i$ with the structure 
\be
\label{smallln}
-ie^{-\theta}\ln\left[\frac{\mu}{2}(z-i(2T-T_0))\right]\;.
\ee
This information enables one to cast the $T/\theta$ problem in the limit
$\theta\to +\infty$ into the form suitable for numerical
analysis. As already discussed, this limit is of primary interest as it
corresponds to semiclassically vanishing number of incoming
particles (this means that the real number of incoming
particles is much smaller than $1/g^2$).
We now proceed to the formulation of the corresponding
equations. 

According to Eq.~(\ref{phichi}), the function $\phi_i$ becomes 
regular in the half-plane $\Im z>T$ when $e^{-\theta}$ tends to zero. 
As for its singularity
in the strip $0<\Im z<T$, it hits the line $\Im z=T$ in the considered
limit $\theta\to +\infty$, 
as is clear already from the general reasoning 
in Footnote~\ref{foot}.  
To demonstrate this explicitly, we write
\be
\label{Ritilde}
\phi_i=i\ln\left[\frac{\mu}{2}(z-iT_0)\right]
-ie^{-\theta}\ln\left[\frac{\mu}{2}(z-i(2T-T_0))\right]
+\tilde{R_i}(z)\;, 
\ee
where $\tilde{R_i}(z)$ is regular both at $iT_0$ and
$i(2T-T_0)$. Equation (\ref{RiT_0'}) then implies
\be
\label{T-T0}
\frac{e^{-\theta}}{T-T_0}=-2\tilde{R_i}'(iT_0)\;.
\ee
This formula should be compared to Eqs.~(\ref{alphaN0}), (\ref{TT0}) 
valid for
the solutions below the critical energy. In that case 
$\tilde{R_i}'(iT_0)=-1/T\neq 0$. 
Let us assume that $\tilde{R_i}'(iT_0)\neq 0$ also for the solutions   
at energies above $E_c$. Then, Eq.~(\ref{T-T0}) implies that 
$T_0$ indeed approaches $ T$ when $e^{-\theta}\to 0$. We conclude that in 
the limit $\theta\to +\infty$ the function $\phi_i$ has 
the form 
\be
\label{phiiN0}
\phi_i=i\ln\left[\frac{\mu}{2}(z-iT)\right]+\tilde{R_i}(z)\;,
\ee
where $\tilde{R_i}(z)$ is regular in the upper half-plane.

As in Sec.~\ref{sec:direct}, it is convenient 
to consider equations for
the functions $\phi_i$ and $\phi_f$ on the real 
axis\footnote{The reader should 
not confuse $x$ which is the real part of the
variable $z$, with the spatial coordinate ${\rm x}$.} 
$z=x\in\mathbb{R}$. 
Again, one should be
careful about the asymptotics  
$x\to\pm\infty$. 
The function $\phi_i$ satisfies the condition (\ref{phiias}).
Taking into account the logarithmic cut of
$\phi_i$ in the strip $0<\Im z<T$ 
(cf. Sec.~\ref{sec:direct}) and conditions (\ref{sol*}), 
one obtains
\bseq
\label{boundreal*}
\begin{align}
\label{boundreal1}
&\phi_i\to 2\pi,~~\phi_f\to 0~~~~~{\rm as} ~~x\to -\infty\;,\\
\label{boundreal2}
&\phi_i\to 0,~~~\;\phi_f\to \pi~~~~~{\rm as}~~x\to +\infty\;.
\end{align}
\eseq
Note that the asymptotics of $\phi_f$ at $x=+\infty$ corresponds to the 
formation
of the sphaleron at the end of the tunneling process.

A convenient expression for the action functional is obtained in the
following way,
\begin{align}
2\Im S&=
2\Im\int_{\cal C}dz (\phi_f\phi_i'-\mu(1-\cos(\phi_i+\phi_f)))\notag\\
&=
-4\pi\Im\tilde{R_i}(iT)+4\pi+
2\Im\int_{-\infty}^{\infty}dx
(\phi_f\phi_i'-\mu(1-\cos(\phi_i+\phi_f)))\;.
\label{actionreal}
\end{align}
In the second line we used Eq.~(\ref{RiT_0}) and took
the limit $T_0\to T$. Integration in the last term 
is performed along the real axis; the first two
terms account for the residue of the integrand in 
the logarithmic singularity of
$\phi_i$. Finally, let us present convenient formulae for
the energy of a solution. One finds two different expressions for 
the energies of the initial and final states; we denote these
energies $E_i$ and $E_f$ respectively\footnote{Evidently, $E_i$ equals
$E_f$ for any solution of the equations of motion.}.
For the final energy it is straightforward to obtain
\be
\label{Efinal}
E_f=2\mu+\int_{-\infty}^{\infty}dz(\phi_f')^2\;,
\ee 
where the first term is due to the presence of the sphaleron in the
final field configuration. 
The analogous expression for the initial energy has the form 
\be
\label{Einitial0}
E_i=\int\limits_{-\infty+iT}^{+\infty+iT}dz(\phi_i')^2\;.
\ee
In the limit $e^{-\theta}\ll 1$ the integral in (\ref{Einitial0}) 
is saturated by the contribution of the
singularity at $z=i(2T-T_0)$. 
One obtains
\be
\label{Einitial}
E_i=2\pi\frac{e^{-\theta}}{T-T_0}=-4\pi\tilde{R_i'}(iT)\;,
\ee
where in the second equality we used Eq.~(\ref{T-T0}).

We have performed numerical solution of the following set of equations
formulated on the real axis:  
differential equation (\ref{boundarynew}) with the 
boundary conditions (\ref{boundreal*}), 
condition of reality of $\phi_f$ (Eq.~(\ref{eqf}))
and analyticity condition
(\ref{phiiN0}). 
Details of our
numerical method are presented in Appendix~\ref{app:B}. Let us mention
here an interesting property of the solutions. 
One notes that the above set of equations is invariant under 
the transformation
\bseq
\label{symm*}
\begin{align}
\label{symm1}
&\phi_i(z) \mapsto 2\pi-\big(\phi_i(-z^*)\big)^*\;,\\
\label{symm2}
&\phi_f(z) \mapsto \pi-\big(\phi_f(-z^*)\big)^*\;.
\end{align} 
\eseq
We find that the tunneling solutions at $E>E_c$ are symmetric with
respect to this transformation. This property distinguishes them
from the solutions at energies lower than critical.

The imaginary part of the action, the energy and suppression exponent are
calculated according to formulae (\ref{actionreal}), (\ref{Efinal}),
(\ref{Einitial}); we use the equality of the initial and final energies as a
cross--check of the precision of numerical calculations. 
Results for the suppression exponent $F(N=0)$ are
presented in Fig.~\ref{fig:1}. They cover the interval $E_c< E < 2.3E_S$
of the region~II. 
Let us stress that our numerical solutions describe tunneling with
exactly {\em zero} semiclassical number of incoming particles. 
The fact that one is able to find such solutions is a peculiarity of
the model.
Unlike in the case of more
complicated systems \cite{Bezrukov:2003er} we 
do not need to perform calculations at finite $N$.

At energies higher than $2.3E_S$ the numerical analysis 
with the method outlined in this subsection becomes problematic. 
The point is
that with the
growth of energy, the parameter $T$ decreases, and the logarithmic
singularity of the solution approaches the 
real axis, according to Eq.~(\ref{phiiN0}). Hence, the finite--difference 
approximation used in numerical calculations
breaks down in the region near $z=0$ at small
$T$. Luckily, as we show in the next subsection, at $\mu T\ll 1$ the
singularity can be
isolated and treated analytically, thus allowing for complete determination
of the solution in the limit $T\to 0$. The energy and suppression exponent
of the limiting solution yield the point $(E_l,F(E_l))$ of the graph
in Fig.~\ref{fig:1}. The interval of energies between $2.3E_S$ and $E_l$
is covered by expanding the solution at these energies in powers of
the small parameter $\mu T$ around the limiting solution, see
Appendix~\ref{app:C}. 

\subsection{Limit $T\to 0$}
\label{sec:limiting}

\begin{figure}[tb]
\begin{center}
\includegraphics[height=7cm,width=10cm]{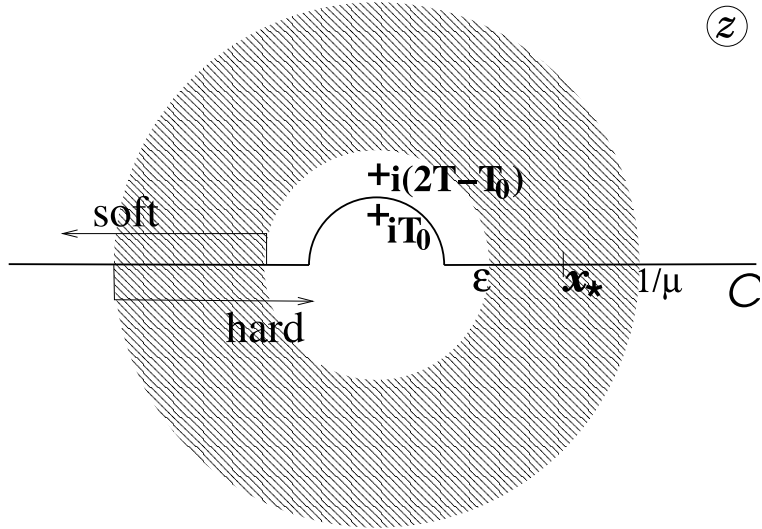}
\end{center}
\caption{``Hard'' and ``soft'' regions of the complex plane;
respective pieces of solution
are glued together in the shaded area. Crosses represent 
the singularities of 
the function $\phi_i$ in the upper half--plane.}
\label{Fig:hardsoft}
\end{figure}

In this subsection we determine the tunneling solution in the limit 
$T\to 0$. We will have to resolve the structure of the singularity of
the solution in the region near the point 
$z=0$, so we keep $T$ small but non-zero,  
$\mu T\ll 1$, and take the limit only in the final formulae for the
energy and suppression exponent. For the same purpose we introduce  
small but
non-vanishing $e^{-\theta}\ll 1$. 
It is convenient to choose the contour ${\cal C}$, entering the 
formula~(\ref{actionnew}) for the action functional,
as shown in Fig.~\ref{Fig:hardsoft}. 
From Eqs.~(\ref{phiias}),
(\ref{sol*}) one obtains that the
asymptotics of the solution on the left part of the 
contour ${\cal C}$ is given by
\be
\label{boundaryC'}
\phi_i\to 0,~~\phi_f\to 0~,~~~{\rm at} ~~x\to -\infty\;,
\ee
while Eq.~(\ref{boundreal2}) remains intact. 
The difference between Eq.~(\ref{boundaryC'})
and Eq.~(\ref{boundreal1}) is explained by the fact that the
contour ${\cal C}$ and the real axis lie on different sides of the
logarithmic cut of the function $\phi_i$.
Correspondingly, 
the symmetry of solutions considered along the contour ${\cal C}$
is 
slightly different from (\ref{symm*}):
instead of (\ref{symm1}) one has
\be
\label{symmC'}
\phi_i(z) \mapsto -\phi_i^*(-z^*)\;.
\ee
We search for solutions invariant under the transformation
(\ref{symmC'}), (\ref{symm2}).

Let us introduce $\varepsilon={\rm max}\{T,e^{-\theta}/\mu\}$.
Our strategy is to separate the complex plane into ``hard'' ($|z|\ll
1/\mu$) and ``soft'' ($|z|\gg \varepsilon$) regions, 
solve equations separately 
in these
regions and glue solutions in their intersection  
$\varepsilon\ll |z|\ll 1/\mu$, see
Fig.~\ref{Fig:hardsoft}. 
Let us start with the ``hard'' region. Inserting
the expression (\ref{Ri}) for $\phi_i$ into Eq.~(\ref{boundarynew})
and neglecting terms of order $O\big((\mu z)^2\big)$, 
one obtains\footnote{It is amusing that Eq.~(\ref{eqhard}) is symmetric with
respect to conformal transformations 
${\phi_{i,f}(z)\mapsto \phi_{i,f}(w(z))-\frac{i}{2}\ln{\frac{dw}{dz}}}$.}
\be
\label{eqhard}
R_i'-\phi_f'=-\frac{i}{z-iT_0}\left(1+e^{i(R_i+\phi_f)}\right)\;.
\ee 
An obvious solution to this equation is
\be
\label{solhard}
R_i=\phi_f=\frac{\pi}{2}\;. 
\ee
The corresponding function $\phi_i$ satisfies the $\theta$-condition
(\ref{phichi}) to the zeroth order in $e^{-\theta}$. However, we saw
in Sec.~\ref{sec:jumps1} that in this approximation the solution is
singular on the contour ${\cal C}$ (cf. Eq.~(\ref{T-T0})). So we need
to find the regularized 
solution which satisfies the $\theta$-conditions to the
first order in $e^{-\theta}$.  
To this end, we linearize Eq.~(\ref{eqhard}):
\be
\label{eqhardlin}
r'-\varphi'=-\frac{r+\varphi}{z-iT_0}\;.
\ee
Here notations 
$r=R_i-\pi/2$, $\varphi=\phi_f-\pi/2$ are introduced. Equation
(\ref{eqhardlin}) can be solved with respect to any of the two functions,
$\varphi$ or $r$. One gets
\begin{align}
\label{varphilin}
&\varphi=-r+(z-iT_0)\int^{z}
\frac{2r'(z_1)}{z_1-iT_0} dz_1 \;,\\
\label{rlin}
&r=\varphi-\frac{2}{z-iT_0}\int_{iT_0}^{z}\varphi(z_1) dz_1\;,
\end{align}
where the integration constant in the 
expression (\ref{rlin}) is chosen in
such a way that the function $r$ is regular at the point
$z=iT_0$. We do not specify the value of the integration constant in
Eq.~(\ref{varphilin}) as this constant is irrelevant for what follows.

Let us investigate 
the singularities of the functions $\varphi$, $r$. 
In the same way as in the previous subsection, one observes that 
the
only singularity of $r$ in the upper half-plane is situated at
$z=i(2T-T_0)$ and has the form (\ref{smallln}) 
(cf. Eq.~(\ref{Ritilde})). 
Then, it follows
from Eq.~(\ref{varphilin}) that the only singularity of $\varphi(z)$
in the upper half-plane is situated at the same point, $z=i(2T-T_0)$
(an apparent 
singularity at $z=iT_0$ is eliminated, as before, by imposing
condition (\ref{RiT_0'}) on the function $r$). 
Reality of $\varphi$ on the
real axis implies now that 
the only other singularity of $\varphi$ is situated at the point
$z=-i(2T-T_0)$, its structure being ``complex conjugate'' to that
of the singularity at $z=i(2T-T_0)$. We determine the form of
these singularities from Eq.~(\ref{varphilin}), single them out
explicitly and expand the remaining entire
function into power series. 
In this way we get the following representation:
\begin{align}
\varphi=&-\frac{e^{-\theta}}{\mu(T-T_0)}\left(\mu (z-iT)
\ln\left[\frac{\mu}{2}(z-i(2T-T_0))\right] 
+\mu(z+iT)
\ln\left[\frac{\mu}{2}(z+i(2T-T_0))\right]\right)\notag\\ 
&-\frac{\pi Te^{-\theta}}{T-T_0}+
A_0\mu z+\sum_{n=1}^{\infty}A_n(\mu z)^{1+2n}\;.
\label{varphisol}
\end{align}
The first term in the second line 
as well as the presence of only odd powers in the power series are fixed by 
the symmetry (\ref{symm2}).
The coefficients in
Eq.~(\ref{varphisol}) are to be determined by matching this
expression 
to the solution
in the ``soft'' region. We will see below that the function $\phi_f$
in the latter region is of order one, and thus all the
coefficients, $\frac{e^{-\theta}}{\mu(T-T_0)}$, $A_0$, $A_n$  
are of order one. So, the contribution 
of the sum in Eq.~(\ref{varphisol}) can be neglected 
in the ``hard'' domain; we omit it in what
follows. Inserting expression (\ref{varphisol}) into Eq.~(\ref{rlin})
one gets: 
\be
\begin{split}
\label{rsol}
r=&-ie^{-\theta}\ln\left[\frac{\mu}{2}(z-i(2T-T_0))\right]+
i\frac{T+T_0}{T-T_0}e^{-\theta}\ln\left[\frac{\mu}{2}(z+i(2T-T_0))\right]\\
&-\frac{4TT_0}{T-T_0}e^{-\theta}\frac{1}{z-iT_0}
\ln\left[1+\frac{z-iT_0}{2iT}\right]-
\frac{e^{-\theta}}{T-T_0}(z+iT_0)+\frac{\pi Te^{-\theta}}{T-T_0}
-iA_0\mu T_0\;.
\end{split}
\ee
In the region $\varepsilon\ll |z|\ll 1/\mu$ the expressions
(\ref{varphisol}), (\ref{rsol}) simplify. In terms of the original
functions $\phi_i$, $\phi_f$, one obtains
\be
\label{phiiint}
\phi_i=i\ln\left[\frac{\mu}{2}(z-i0)\right]
+\frac{\pi}{2}-\frac{e^{-\theta}}{T-T_0}\;z~,~~~~~~~
\phi_f=\frac{\pi}{2}-\frac{e^{-\theta}}{T-T_0}
\;z\ln\left(\frac{\mu^2z^2}{4}\right)+A_0\mu z\;,
\ee
where terms of order $O(\mu T)$, $O(e^{-\theta})$,
$O\big((\mu z)^2\big)$ have been neglected. 

We now turn to the ``soft'' region. 
Here the contour ${\cal C}$ coincides with the real
axis. 
Introducing the real and imaginary parts of the incoming wave,
$\phi_i(x)=a(x)+ib(x)$, one rewrites Eq.~(\ref{boundarynew}) as a 
set of two real equations, 
\begin{align}
\label{bprime}
&b'=\mu\sh{b}\cos{u}\;,\\
\label{uprime}
&u'=2a'-\mu\ch{b}\sin{u}\;,
\end{align}
where $u(x)=a(x)+\phi_f(x)$.
In the ``soft'' region one can neglect the regularizing parameters 
$\mu T$, $e^{-\theta}$.  
[Solutions at non-zero $T$ are considered in Appendix~\ref{app:C}.]
Then, the function $\phi_i$ is regular in the upper
half-plane of the ``soft'' region, 
and the derivatives of its real and imaginary parts are related by 
an analog of the Cauchy formula,
\be
\label{aprime}
a'(x)=\frac{1}{\pi x} {\rm V.P.}\!\!\int_{-\infty}^{\infty}
dx_1 \frac{x_1b'(x_1)}{x_1-x}\;,
\ee 
where the integral is understood in the sense of principal
value. Equations (\ref{bprime}), (\ref{uprime}), (\ref{aprime}) are 
supplemented by boundary conditions at $x=0$ 
which originate from matching with the ``hard'' part of
solution. From  
the expressions (\ref{phiiint}) one obtains
\be
\label{x0}
b\to\ln{\left(\frac{\mu}{2}x\right)}~,~~~
a\to\frac{\pi}{2}~,~~~u\to\pi~,~~~~{\rm at}~~~x\to+0\;.
\ee 
The conditions at $x\to -0$ can be reconstructed from the requirement of
symmetry with respect to (\ref{symmC'}), (\ref{symm2}).
For completeness we write down the 
boundary
conditions
(\ref{boundreal2}), (\ref{boundaryC'}) explicitly:
\be
\label{xpminfty*}
a\to 0~,~~~~b\to 0~,~~~
u\to\pi~,~~~~~{\rm at}~~x\to +\infty\;,
\ee
while the symmetry transformations (\ref{symm2}), (\ref{symmC'}) are:
\be
\label{symmetry}
b(x)\mapsto b(-x)~,~~~a(x)\mapsto -a(-x)~,~~~u(x)\mapsto\pi-u(-x)\;.
\ee
Due to the symmetry (\ref{symmetry}), it is sufficient to consider 
half of the real axis, and we assume $x>0$ in what follows.  
Given the function $u$,
equation (\ref{bprime}) can be integrated explicitly:
\be
\label{bexplicit}
b(x)=\ln{\tanh\left(-\frac{\mu}{2}\int_0^x\cos{u(x_1)}dx_1\right)}\;,
\ee
where the condition (\ref{x0}) is imposed on $b(x)$ at the 
origin\footnote{Note that due to the asymptotics (\ref{xpminfty*}) 
of $u(x)$ one has
$b\propto e^{-\mu x}$ at large values of $x$. Such behavior indicates
that at $t\to +\infty$ the imaginary part of the solution describes
the evolution along the negative mode of the sphaleron. 
This is a generic feature
of tunneling solutions describing jumps onto the sphaleron 
 \cite{Bezrukov:2003tg}.}.
Equations \eqref{bexplicit}, (\ref{aprime}), (\ref{uprime}) form a
complete system.

\begin{figure}[tb]
\begin{center}
\includegraphics[height=5cm,width=0.45\textwidth]{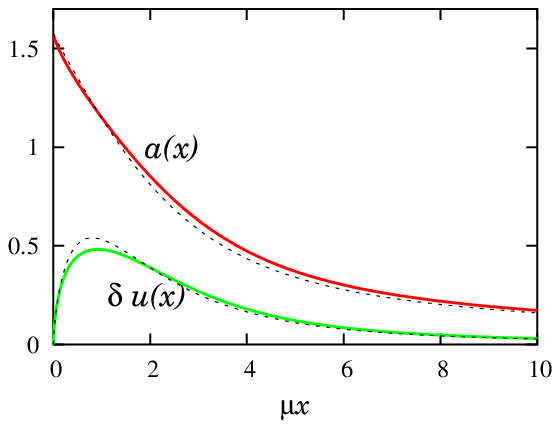}
\includegraphics[height=5cm,width=0.45\textwidth]{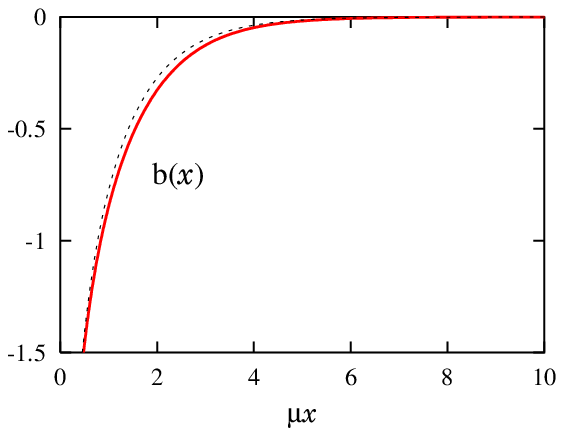}
\end{center}
\caption{Limiting solution in the ``soft'' region. Dashed lines
represent linear approximation in $\delta u$.}
\label{Fig:abu}
\end{figure}

One obtains a fairly good approximation by
linearizing Eqs.~(\ref{bexplicit}), (\ref{uprime}) 
with respect to $\delta u=u-\pi$. 
Equation (\ref{bexplicit}) takes the form
\[
b=\ln{\tanh{\left(\frac{\mu}{2}x\right)}}\;.
\] 
Substitution of this expression into Eq.~(\ref{aprime}) yields
\[
a'=-\frac{\mu}{\pi}\left(\beta\left(1+i\frac{\mu}{\pi}x\right)
+\beta\left(1-i\frac{\mu}{\pi}x\right)\right)\;,
\]
where 
\[
\beta(x)=\frac{d}{dx}\left(\ln{\left[\Gamma\left(\frac{x+1}{2}\right)\right]}-
\ln{\left[\Gamma\left(\frac{x}{2}\right)\right]}\right)\;.
\]
Now, $\delta u$ can be found from Eq.~(\ref{uprime}). The corresponding
analytic
expression is not illuminating, and we do not present it here. 
The linear approximation is improved by 
numerical iterations. Each cycle consists of solving 
Eqs.~(\ref{bexplicit}), (\ref{aprime}), (\ref{uprime}) thus
determining  
functions $b$, $a$, $\delta u$ one
after another, starting from the
approximation for $\delta u$ obtained in the previous cycle.
After $30$ iterations one obtains numerical solution with precision of
order $10^{-6}$. 
The resulting functions $a$, $b$, $\delta u$ 
together with those obtained in the
linear approximation
are plotted in Fig.~\ref{Fig:abu}.

Let us consider the behavior of 
the solution at $x\to +0$. 
Equation~(\ref{uprime}) allows for the following behavior: 
\be
\label{nearzero}
a=\frac{\pi}{2}-A\mu x+O(x^2)~,~~~~~~~
\phi_f=\frac{\pi}{2}-2A\mu x\ln{\left(\frac{\mu x}{2}\right)}+
B\mu x+O(x^2)\;.
\ee
We observe that both the
solution
in the linear approximation and the full
numerical solution exhibit this behaviour. Numerically,
$A=0.845$, $B=-2.90$. 
Matching these expressions to the asymptotics (\ref{phiiint})
of the ``hard'' part of the solution, one obtains
\be
\label{matching}
\frac{e^{-\theta}}{T-T_0}=A\mu~,~~~A_0=B\;.
\ee
Thus, the limiting 
solution is completely determined.

A comment is in order. Strictly speaking, in the limit
$T, e^{-\theta}\to 0$ the ``soft'' region covers the whole complex
plane and one is left with the ``soft'' part of the solution
only. However, 
the solution is then singular at $z=0$, and it is impossible to
calculate its energy and action. So the term ``limiting solution''
implies not only the form of the solution in the ``soft'' region but
also the appropriate resolution of the singularity, 
Eqs.~(\ref{varphisol}), (\ref{rsol}).

The next step is to evaluate 
the imaginary part of the action functional
on the constructed solution. It is natural to 
separate it into the sum of contributions of the ``hard'' and
``soft'' regions. 
Neglecting terms of order $\mu T$, $e^{-\theta}$ we have
\[
2\Im S_{hard}=2\Im\int\limits_{C, |z|<x_*}
\left(\frac{i\pi}{2(z-iT_0)}-\mu-\frac{1}{z-iT_0}\right)~dz =2\pi\;.
\] 
Here $x_*$ is some number in the range $\varepsilon\ll x_*\ll 1/\mu$.
For the ``soft'' contribution we write
\be
\label{Ssoftcalc}
\begin{split}
2\Im S_{soft}&=2\int_{|x|>x_*} (\phi_f b'-\mu\sh{b}\sin{u})~dx\\
&=2\pi b(x_*)+4\int_{x_*}^{\infty}(\phi_f-\tg{u})b'~dx
=4\int_{0}^{\infty}\left(\frac{u'}{\cos^2 u}-\phi_f'\right)b~dx\;,
\end{split}
\ee
where in
passing from the first to the second line we used 
Eq.~(\ref{bprime}) and symmetry with respect to 
(\ref{symmetry}), while in the last expression terms vanishing
in the limit 
$x_*\to 0$ are neglected.
A good estimate for $2\Im S_{soft}$ is obtained in the linear
approximation:
\[
2\Im S_{soft}^{lin}=4\int_0^{\infty}a'b~dx
=4\pi\left(-6\zeta'(-1)-\frac{\ln{2}}{6}-\frac{\ln{\pi}}{2}\right)=3.83\;,
\]
where
integration is performed using Eqs.~8.371.2, 3.951.14 from
Ref.~\cite{Ryzhik}. 
The calculation of the integral (\ref{Ssoftcalc}) 
for the full numerical solution yields 
$2\Im S_{soft}=3.98$. Note that this value is comparable
to the contribution $2\pi$ coming from the ``hard'' core.

An important quantity is the energy of the limiting solution. Again, 
we calculate
separately the energies of the initial and final states.
While formula (\ref{Efinal}) for the final energy is directly applicable
in the limit $T\to 0$, the analog of Eq.~(\ref{Einitial}) for the initial
energy is
\[
E_i=2\pi\frac{e^{-\theta}}{T-T_0}=2\pi A\mu\;,
\]    
where $A$ is defined in Eqs.~(\ref{nearzero}). For our
numerical solution $E_f=E_i$ with good 
accuracy, this may be viewed
as a confirmation of the consistency of the numerical method; the obtained
value is
\be
\label{Elimit}
E_l=2.65E_S\;.
\ee
At first sight, the result (\ref{Elimit}) is puzzling. Indeed, 
one could expect to cover the whole range of energies up to infinity
by sending the parameter $T$ to zero. 
However
from Eq.~(\ref{Elimit}) we learn that this is not the case:
$T/\theta$-problem does not have any solutions with  
energies higher than $E_l$.
On the other hand, the suppression exponent is still
non-vanishing at the limiting energy, 
\be
\label{Flimit}
F(N=0,E=E_l)=10.27\;.
\ee
A question is then, how tunneling occurs
at energies above $E_l$, 
and what is the corresponding tunneling
probability. 

The answer is suggested by an observation that in all the
formulae describing the limiting solution, even in those related to the
``hard'' core, one can set $T=0$ (it is sufficient to keep 
$e^{-\theta}$ non-zero to
resolve the singularity of the solution). 
As discussed at the end of
Sec.~\ref{sec:general}, the solution obtained in this way belongs to
the class of real--time instantons. Thus, it gives the
maximum probability for tunneling from states with semiclassically
vanishing number of
particles. Consequently, the suppression exponent $F(N=0,E>E_l)$ is not
smaller than $F(N=0,E=E_l)$. It cannot be larger either, because one can
always imagine a tunneling process which 
has the same exponential suppression as the
process at the limiting energy. Namely, 
incoming particle(s) can release the energy excess $(E-E_l)$ by
the perturbative emission of a few other particles at the beginning of the
process, so that subsequent tunneling occurs effectively at the energy
$E_l$.  
From this physical reasoning one
concludes that the suppression exponent stays constant at energies higher
than $E_l$ (cf. Ref.~\cite{Voloshin:1993dk}). Remarkably,
the semiclassical treatment is still possible at these energies. We 
return to this issue in Sec.~\ref{sec:emission}. Now, let us turn 
to the real--time instantons which, for their
novelty, deserve a somewhat detailed study.

\subsection{Real-time instantons}
\label{sec:real-time}

Let us recall that the real-time instantons 
are solutions of the $T/\theta$-problem with $T=0$ and finite 
$\theta$.
In the model under consideration these solutions can be obtained
by an iterative method which is a slight modification of that
used in the previous
subsection. As before, one introduces the functions $a$, $b$, $u$
satisfying Eqs.~(\ref{bprime}), (\ref{uprime}). 
Note that, unlike in the case of Sec.~\ref{sec:limiting}, all the
functions are supposed to be regular on the entire real axis,  
including the
point $x=0$. 
\begin{figure}[!ht]
\begin{center}
\includegraphics[height=6cm,width=0.8\textwidth]{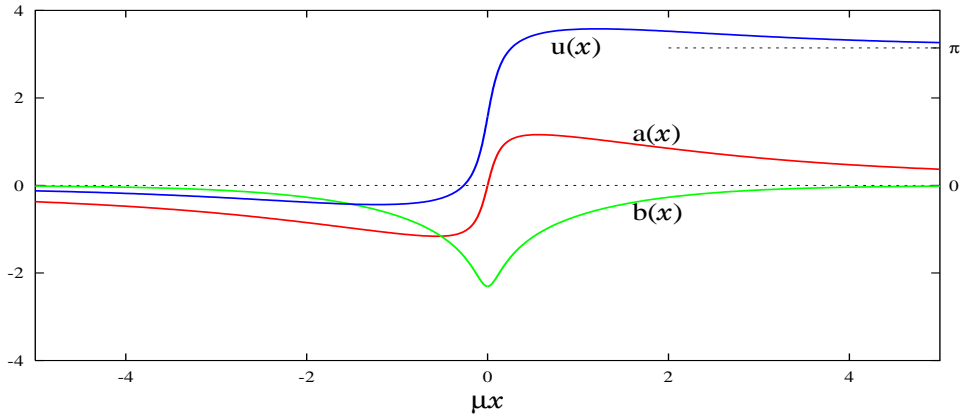}
\end{center}
\caption{Example of real-time instanton configuration with 
$\theta=2.04$, $N=2.3$.}
\label{Fig:real-time}
\end{figure} 
Requirements of regularity together with symmetry under
the transformations (\ref{symmetry}) lead to the following conditions
at $x=0$,  
\be
\label{x0new}
b'(0)=0~,~~~a(0)=0~,~~~u(0)=\pi/2\;.
\ee
Formula (\ref{aprime}) is not directly applicable in the
case of finite $\theta$. 
An appropriate modification comes about when one notices 
that a direct analog of Eq.~(\ref{aprime}) relates the real
$\tilde a$
and imaginary $\tilde b$ parts of the function $\chi$
introduced in Eqs.~(\ref{phichi}), (\ref{chi}). 
Using Eq.~(\ref{phichi}), one expresses the functions $a$, $b$ in terms of
$\tilde a$, $\tilde b$
and gets, instead of Eq.~(\ref{aprime}),
\be
\label{aprimenew}
a'(x)=\frac{1+e^{-\theta}}{1-e^{-\theta}}\cdot
\frac{1}{\pi x} {\rm V.P.}\!\!\int_{-\infty}^{\infty}
dx_1 \frac{x_1b'(x_1)}{x_1-x}\;.
\ee
Finally, Eq.~(\ref{bprime}) can be integrated explicitly in the
present case as well, but now the
integration constant is arbitrary,
\be
\label{bnew}
b(x)=\ln{\tanh
\left(-\frac{\mu}{2}\int_0^x\cos{u(x_1)}dx_1+\varkappa\right)} \;.
\ee
We obtain numerical solutions by iterating 
Eqs.~(\ref{bnew}), (\ref{aprimenew}), (\ref{uprime}) 
in the indicated order.  The value of  $\varkappa$
is determined at each cycle of iterations 
from the requirement that the function $u$ satisfies
the boundary conditions (\ref{x0new}), (\ref{xpminfty*}).
An example of real--time instanton configuration is shown in
Fig.~\ref{Fig:real-time}.

\begin{figure}[t]
\begin{center}
\includegraphics[height=6cm,width=0.45\textwidth]{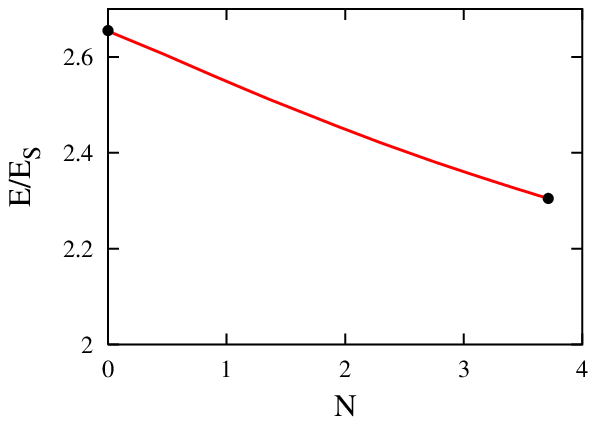}
\includegraphics[height=6cm,width=0.45\textwidth]{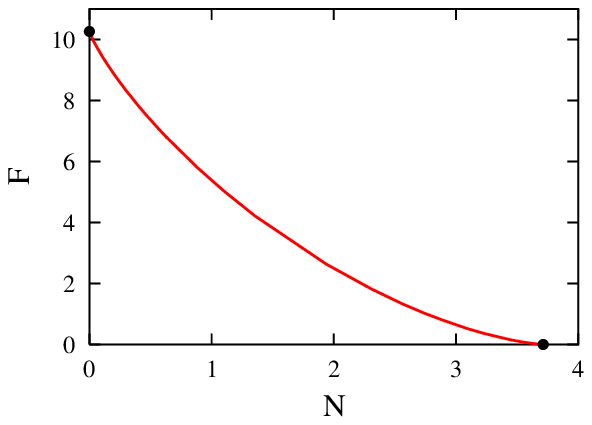}
\end{center}
\caption{Energy and suppression exponent versus the number of incoming
particles in the case of real-time instantons.} 
\label{Fig:rtEF}
\end{figure}

It is straightforward to 
obtain the energy and imaginary part of the action from 
Eqs.~(\ref{Efinal}) and (\ref{actionnew}). As for the number of incoming
particles, in the case of real--time instantons it can be expressed
in a particularly convenient form,
\[
N=-\frac{2}{\sh{\theta}}\int_{-\infty}^{+\infty} a'b~dx\;,
\]
which can be obtained from Eq.~(\ref{EN2}) in a
straightforward way.
The energy and suppression
exponent as functions of the initial number of particles
are shown in Fig.~\ref{Fig:rtEF}. 
The curves start at $N=0$ from the values
corresponding to the limiting solution and end up at $N=3.7$ where the
suppression exponent is zero. We see that the branch of
real-time instantons interpolates between the limiting solution and
the region of classically allowed transitions.  

\subsection{Tunneling above the limiting energy}
\label{sec:emission}

In this subsection we follow an approach which is 
somewhat different from that
used in the rest of the paper. Instead of considering the inclusive
probability (\ref{eq:PEN}), we study tunneling from a
{\em single} initial state with given number of particles $N$ and energy
$E$. The energy is assumed to be 
higher than the energy of the real--time instanton
corresponding to the same initial number of particles, $E>E^{(rt)}(N)$. 
Let us demonstrate that the suppression exponent of tunneling from an
appropriately
chosen initial state coincides with
that of the real--time instanton.

Using the approach of
Ref.~\cite{Rubakov:1992az}, 
one can show that any field configuration which
satisfies Eqs.~(\ref{eq:eqa}), (\ref{eq:eqaa}), (\ref{eq:eqb}) 
(but not necessarily Eq.~(\ref{eq:eqc}))
in the
complex time plane determines the suppression exponent for the tunneling
probability of the form
\be
\label{PENa}
\tilde{\cal P}(E,N;a)=
\sum_f\frac{|\bra{f}\hat{\cal S}\hat{P}_E\hat{P}_N\ket{a}|^2}{
\bra{a}\hat{P}_E\hat{P}_N\ket{a}}\;.
\ee
Here summation is performed over the final states only, the initial state 
$\hat{P}_E\hat{P}_N\ket{a}$ is a coherent state projected onto the
subspace with fixed energy and number of particles. The
coherent state corresponding to a given solution is defined by 
\be
\hat{a}_k\ket{a}=f_k\ket{a}\;,
\ee
where $f_k$ are the positive frequency components of 
the solution $\phi(t,{\rm x})$ at $t\to -\infty$,
\[
\phi(t,{\rm x})\Big|_{t\to-\infty} = 
\frac{1}{(2\pi)^{1/2}}\int \frac{dk}{\sqrt{2\omega_k}}\left(
f_k \mathrm{e}^{-i\omega_k t+ik{\rm x}} + 
g_k^*\mathrm{e}^{i\omega_k t-ik{\rm x}}\right)\;.
\]
Note that unlike in Eq.~(\ref{philinear}) we now 
consider the asymptotics along the real $t$-axis. 
The suppression exponent for the 
probability (\ref{PENa}) is still given by Eq.~(\ref{F})
where the parameters $T$ and $\theta$ are now determined from the
following equations,
\be
\label{ENa}
E=\int dk\omega_k f_kf_k^*\e^{2\omega_kT+\theta}~,~~~~
N=\int dk f_kf_k^*\e^{2\omega_kT+\theta}\;.
\ee
One can check that, as before,
$E$ coincides with the actual energy of the solution, and 
the expressions (\ref{EN}) for the energy and number of incoming particles
are applicable.  
Note that in general 
the average energy and number of particles in the
state $\ket{a}$ are different from $E$ and $N$.

We consider the following configurations which differ from 
the real--time instanton by a small perturbation:
\be
\label{excit*}
\phi_{i,f}(z)=\phi_{i,f}^{(rt)}(z)+\alpha\psi(z/\beta)\;,\\
\ee
where the perturbation is represented by the function $\psi(z)$ which
is real 
on the real axis and falls off fast enough at $x\to\pm\infty$. 
When the parameters $\alpha$, $\beta$ are small, $\beta\ll\alpha\ll 1$,
the configuration (\ref{excit*}) is an approximate solution of
Eq.~(\ref{boundarynew}). Evaluation of the energy and
number of incoming particles using Eqs.~(\ref{Efinal}),
(\ref{EN2}) yields
\begin{align}
&E=E^{(rt)}+\frac{\alpha^2}{\beta}
\int dx(\psi'(x))^2 
+O(\alpha)\;,\notag\\
&N=N^{(rt)}+O(\alpha^2)\;\notag.
\end{align} 
These formulae
demonstrate that in the limit 
\[
\alpha,\beta\to 0~,~~~ ~ 
\alpha^2/\beta~~{\rm fixed}
\] 
the solution (\ref{excit*}) describes tunneling
from a state which differs from the initial state of the real--time
instanton by addition of a few particles carrying a finite 
amount of energy. Comparison of the
structure of the initial and final waves shows that these particles 
scatter elastically from the boundary ${\rm x}=0$ and do not affect
the tunneling process.
It is straightforward to evaluate the imaginary part of the action of
the 
solution (\ref{excit*}) and determine the parameters $T$ and $\theta$
from the formulae~(\ref{ENa}). In this way one finds that 
the suppression exponent of the solution
(\ref{excit*}) coincides with that
of the real--time instanton
up to terms~$O(\alpha)$.

As before, the suppression exponent of collision--induced tunneling is
obtained in the limit $N\to 0$. One concludes that it 
is equal to that of the limiting solution at
all energies above~$E_l$.

\section{Concluding remarks}
\label{sec:discussion}

The results
obtained in this paper show that in the model (\ref{eq:S}) the soliton
production due to collision of a particle with the boundary is always
exponentially suppressed. This may be viewed as an evidence in favour of 
the exponential suppression of collision--induced tunneling at
all energies in other theories.   
We have identified the mechanism of
suppression at high energies of colliding particles. It is a
consequence of substantial rearrangement the system has to undergo in
the course of transition. Moreover, this rearrangement occurs in an
optimal way at a certain limiting energy $E_l$ where the suppression
is the weakest. At higher energies the system emits the excess of
energy in the form of particles which do not interfere with the tunneling
process,
and subsequent transition occurs with the energy $E_l$. The consequence of
the latter phenomenon is that the suppression exponent stays constant all the
way from $E=E_l$ to $E=\infty$. It is natural to conjecture that
this mechanism of suppression is generic for collision--induced
tunneling in quantum field theory.

As a by-product we have found that there 
is a novel type of tunneling
configurations 
which we call real--time instantons. These are solutions to the
$T/\theta$-problem with the parameter $T$ set equal to zero. From the physical 
point of view, real--time instantons 
provide maximum probability for tunneling from
initial states with given number $N$ of incoming particles. In the plane
of parameters $(E,N)$, the line of real--time instantons interpolates
between the region of classically allowed transitions and
the point $(E=E_l,N=0)$. The suppression exponent calculated for 
a real--time
instanton $F_{rt}(N)$ sets a lower bound on the suppression
exponent $F(E)$ for the tunneling induced by one or a few incoming particles 
at all energies. 
Modulo standard
assumptions, $F_{rt}(N)$ in the limit $N\to 0$ coincides with $F(E)$ at
energies higher than $E_l$. 
Thus,
one can use real--time instantons to determine the value of the limiting
energy $E_l$ and the suppression exponent of collision--induced tunneling 
at energies higher than $E_l$
in various models.

Finally, let us recall that the model investigated in this
paper describes various condensed matter systems. It would be
interesting to work out implications of our results for these systems.

\paragraph*{Acknowledgements}
We are indebted to S.~Dubovsky for bringing to our attention the model
considered in this paper and for 
collaboration at early stage of the work. We thank
V.~Rubakov and F.~Bezrukov for their encouraging interest and helpful
suggestions.
We are grateful to S.~Demidov, D.~Gorbunov, M.~Libanov, E.~Nugaev,
C.~Rebbi, A.~Ringwald, G.~Rubtsov, S.~Troitsky for fruitful
discussions. 
This work has been supported in part by Russian Foundation for Basic
Research, grant 02-02-17398, grant of the President
of Russian Federation NS-2184.2003.2 and personal felloships of the ``Dynasty''
foundation (awarded by the Scientific board of ICFPM).
The work of D.~L. has been supported by CRDF award RP1-2364-MO-02 and 
INTAS grant YS 03-55-2362. 
S.~S. is grateful to DESY Theory group in Hamburg for hospitatlity
during his visit. 

\appendix
\section{Periodic instantons at non-zero bulk mass}
\label{app:A0}
Let us consider the effect of non-zero bulk mass $m$ on the periodic
instantons found in Sec.~\ref{sec:periodic}. We assume that the mass
is small, $mT\ll 1$. An approximate periodic instanton 
$\phi_m(t,{\rm x})$ in the massive case is constructed in the
following way. In the region $|t|\ll 1/m$, one has
\be
\label{m0}
\phi_m(t,{\rm x})=\phi_0(t,{\rm x})\e^{-m{\rm x}}\;,
\ee  
where $\phi_0(t,{\rm x})$ is the periodic instanton solution  in the
massless case given by Eqs.~(\ref{decomp}),
(\ref{PIsecond}). 
It is straightforward to check that 
the configuration (\ref{m0}) satisfies the field equations 
(\ref{eq:eqa}), (\ref{eq:eqaa}) up to terms of order $O(mT)$. At
$|t|\gg T$ the value of the field at the boundary is 
frozen to $0$ and $2\pi$ in the
parts $A$ and $D$ of the contour $ABCD$,
respectively (see Fig.~\ref{fig:2}). 
Let us concentrate on the initial part of the
contour; we set $t=t'+iT$. In the region $T\ll |t'|\ll 1/m$, the 
configuration (\ref{m0}) should be matched to the solution of the free
massive equation, which has the form
(\ref{philinear}). Making use of Eq.~(\ref{inpacket}), one obtains the
Fourier transform of the function $\phi_m$:
\[
\phi_m(t',{\rm x})=\int_{-\infty}^{\infty}\frac{dk}{k-im}\cdot
\frac{-i\e^{ikx_0}}{2\ch\left(\frac{kT}{2}\right)}
\e^{ik(t'+{\rm x})}~~,~~~~~~T\ll |t'|\ll \frac{1}{m}\;,
\]
where we omitted terms negligible in the limit $m\to 0$. Comparing
this expression to Eq.~(\ref{philinear}) we get,
\begin{align}
&f_k=\theta(-k)
\frac{-i\sqrt{\pi\omega_k}\e^{ikx_0}}{(k-im)\;
\ch\left(\frac{kT}{2}\right)}\;,
\notag\\
&g^*_k=\theta(-k)
\frac{-i\sqrt{\pi\omega_k}\e^{-ikx_0}}{(-k-im)\;
\ch\left(\frac{kT}{2}\right)}\;.
\notag
\end{align}
Now, it is straightforward to evaluate the energy and the number of
incoming particles using Eqs.~(\ref{EN}). The energy is given by the
same formula as in the massless case, Eq.~(\ref{PIenergy}), 
while the initial
number of particles is
\[
N=\int_0^{\infty}\frac{\pi}{\sqrt{k^2+m^2}\;
\ch^2\!\!\left(\frac{kT}{2}\right)}dk
=-\pi\ln(mT)+O(1)\;.
\] 
We see that the small bulk mass does not affect the energy of the
periodic instanton. On the other hand, it regularizes the number of
incoming particles which diverges logarithmically in the limit $m\to
0$. To find the effect of the mass on the value of the action,
one performs integration by parts in expression (\ref{action1}) and
obtains
\[
S=\int dt\left.\left(\frac{1}{2}\phi\d_{\rm x}\phi-
\mu(1-\cos{\phi})\right)\right|_{{\rm x}=0}\;.
\]
From the above consideration it follows that
the integrand is non-vanishing only at $t\sim T\ll 1/m$. 
So, one substitutes the expression (\ref{m0}) into this formula and finds
that the value of the action differs from that in
the massles case by terms 
of order $O(mT)$. We conclude that the influence of the small mass on 
the suppression
exponent of tunneling described by the periodic instanton is small when
the energy of the process is much higher than the mass.

\section{Analytic proof of existence of the critical energy}
\label{app:A}
We are going to prove the following statements concerning Eq.~(\ref{ueq}):\\
a) if $\lambda > 4$, a solution of Eq. (\ref{ueq}) with
asymptotics (\ref{ubound*}) exists;\\
b) if $\lambda <\pi/2$, there is {\em no} solution of Eq. (\ref{ueq}) with
asymptotics (\ref{ubound*}). 

\begin{figure}[htb]
\begin{center}
\includegraphics[height=6cm,width=11.cm]{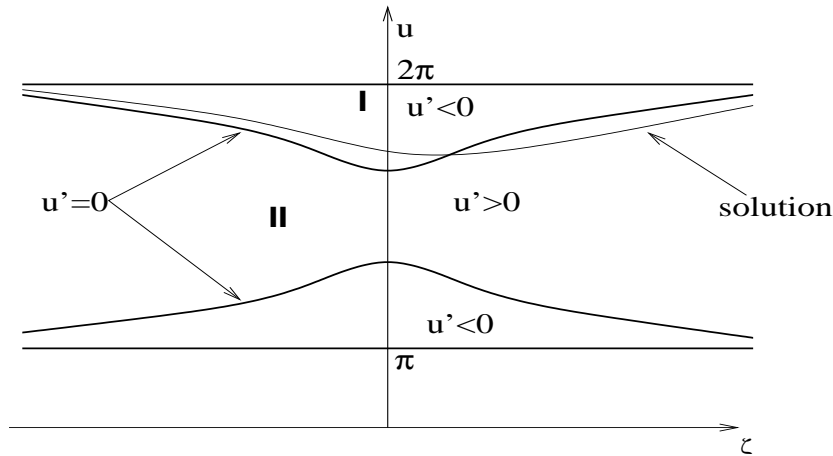}
\end{center}
\caption{Regions of positive and negative $u'$ and qualitative form of
the solution of Eq.(\ref{ueq}) in the case $\lambda>4$.}
\label{lam4}
\end{figure}

We start with the statement (a). At $\lambda>4$
the lines defined by equation
\be
\frac{4}{1+\zeta^2}+\lambda\sin{u}=0
\label{u'0}
\ee 
separate the band $\pi <u<2\pi$ of the
$(\zeta,u)$ plane into three regions, where the derivative of solution of
Eq.~(\ref{ueq}) has different signs, see Fig.\ref{lam4}. 
The condition (\ref{ubound1}) picks up a single solution
of Eq.(\ref{ueq}).
The integral curve of this solution starts from the region I 
at $\zeta=-\infty$. As $\zeta$ increases, it crosses
the line $u'=0$ and comes into the region II where $u'>0$. 
Once in the region II, the integral curve cannot 
leave it.  
Thus, $u'$ stays positive all the way to $\zeta=+\infty$,
and the integral curve gets attracted to the asymptotics $2\pi$ when
$\zeta\to +\infty$. 

\begin{figure}[htb]
\begin{center}
\includegraphics[height=6cm,width=11.cm]{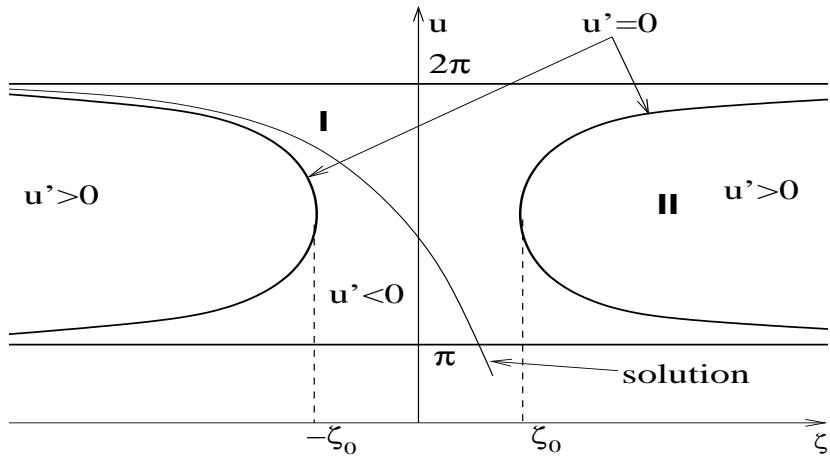}
\end{center}
\caption{Regions of positive and negative $u'$ and qualitative form of
the solution of Eq.(\ref{ueq}) in the case $\lambda< \pi/2$.}
\label{lam2}
\end{figure}

We now proceed to the proof of the statement (b). At $\lambda<\pi/2$, 
the form of the lines (\ref{u'0}) is different, see
Fig.~\ref{lam2}. The tips of these lines are situated at 
$\zeta=\pm \zeta_0$,
where 
\[
\zeta_0=\sqrt{4/\lambda-1}\;.
\]
At $\zeta<-\zeta_0$ the integral curve 
of the solution lies in the region I where $u'<0$. 
Let us show that it 
crosses the line $u=\pi$ from above within the
interval $-\zeta_0<\zeta<\zeta_0$. Indeed, Eq.~(\ref{ueq}) implies,
\[
u'\leq\lambda-\frac{4}{1+\zeta^2}\;.
\]
Integration of the latter inequality yields
\[
u(\zeta_0)\leq u(-\zeta_0)+2\lambda\zeta_0-8\arctg{\zeta_0}<\pi\;.
\]
Once the integral curve comes below the line $u=\pi$, it cannot reach
the region II where $u'>0$ (see Fig.~\ref{lam2}). Thus it does not
approach the asymptotics $2\pi$ at $\zeta\to +\infty$.

The above analysis necessitates the existence of the critical value
$\pi/2<\lambda_c<4$ at which the behavior of the solution changes
qualitatively. We have obtained the value of $\lambda_c$ 
numerically, see Eq.~(\ref{lambdac}).

\section{Numerical method at $E>E_c$}
\label{app:B}
In this Appendix we formulate a discretized version of the 
problem (\ref{boundarynew}), (\ref{eqf}), (\ref{phiiN0}), 
(\ref{boundreal*}). 
When searching for an appropriate numerical 
method, one encounters two major difficulties. 
First, since all the relevant solutions 
at $E>E_c$ are close to the sphaleron configuration 
at $t\to +\infty$, they display an instability similar 
to that of the sphaleron\footnote{Instability of 
this kind is a generic feature of tunneling solutions 
at energies higher than critical, 
see~\cite{Bezrukov:2003tg,Kuznetsov:1997az,Bezrukov:2003er}. 
A general method of dealing with this instability is
proposed in~\cite{Bezrukov:2003tg}.}. 
To overcome this difficulty, we use specifics of 
the system in hand. As already mentioned, equations
(\ref{boundarynew}), (\ref{eqf}), (\ref{phiiN0}) and boundary 
conditions (\ref{boundreal*}) are 
invariant under the transformation (\ref{symm*}).
So, it is natural to look for solutions symmetric with respect 
to this transformation. Once the symmetry requirement is imposed,
the set of considered configurations is restricted to those close to
the sphaleron at $t\to +\infty$, and solutions become stable.

The second numerical challenge is  
the slow falloff of the solutions 
at $x\to\pm\infty$.  Equation (\ref{boundarynew}) allows for solutions 
with asymptotic behavior $1/x$ at $x\to\pm\infty$, and numerical
analysis confirms that this is indeed the case.
We overcome this problem by introducing a non--uniform lattice
with link size growing when $x\to\pm\infty$. This is achieved by
performing  
conformal transformation of the complex $z$--plane,
\begin{equation}
\label{conftrans}
w = \frac{z-iT}{z+iT}\;,
\end{equation}
which maps the real axis onto the the unit circle 
$w = \mathrm{e}^{i\alpha}$. 
In particular, the asymptotic regions $x\to -\infty$ and 
$x\to +\infty$ correspond to the vicinities 
of the points $\alpha = 0$ and 
$\alpha= 2\pi$, respectively.  
It is clear that the uniform lattice 
in angle $\alpha$ produces 
discretization of the required type of the real $z$--axis. 

We proceed by reformulating the problem on the 
circle $|w| = 1$.
First, it is convenient to introduce the function
\[
S_i(z) = \phi_i(z) - i\ln\left(\frac{z-iT}{z+iT}\right)\;.
\]
Due to Eq. (\ref{phiiN0}), $S_i(z)$ is regular in the upper 
$z$--plane. Thus, after
the conformal transformation~\eqref{conftrans} it is regular 
inside the unit 
circle $|w|<1$, and the Taylor series
\begin{equation}
\nonumber
S_i(w) = i\sum\limits_{k=0}^{\infty} S_k w^k
\end{equation}
converges everywhere inside this circle. For 
$w = \mathrm{e}^{i\alpha}$ we obtain
\begin{subequations}
\label{appB:eq}
\begin{equation}
\label{fourSi}
S_i(\alpha) = i\sum\limits_{k=0}^{\infty}S_k \mathrm{e}^{ik\alpha}\;.
\end{equation}
The requirement that the Fourier series of the function $S_i$ has the 
form~\eqref{fourSi} 
is equivalent to the condition (\ref{phiiN0}). 
Equation (\ref{boundarynew}) in new terms reads
\begin{equation}
\label{eqphi}
\frac{d}{d\alpha}\big[S_i - \phi_f\big] = 1 + 
\frac{T\mu}{2\sin^2(\alpha/2)}\sin(S_i + \phi_f - \alpha)\;,
\end{equation}
where the function $\phi_f$ is real due to Eq. (\ref{eqf}).
The reformulation of the boundary conditions (\ref{boundreal*}) 
is straightforward. 
However, all but one of the conditions
(\ref{boundreal*}) are
redundant. Indeed, if the conditions at $x\to -\infty$ ($\alpha\to 0$) are
satisfied, requirements at $x\to +\infty$ ($\alpha\to 2\pi$) hold 
automatically
due to the symmetry (\ref{symm*}).
Moreover, the condition on the
function $S_i$ at $\alpha\to 0$
is satisfied automatically due to 
Eq.~\eqref{eqphi}. The remaining condition implies
\begin{equation}
\label{bc}
\phi_f \to 0 \;\;\;\mathrm{as} \;\;\;\alpha\to 0\;. 
\end{equation}
\end{subequations}
One notes that the symmetry (\ref{symm*}) 
relates the points $\alpha$ and $2\pi-\alpha$
of the circle ${|w| = 1}$. Therefore, Eq.~\eqref{eqphi} 
can be regarded as a differential equation
on the orbifold $\alpha\in [0,\pi]$, with the condition~\eqref{bc} 
imposed at the 
orbifold boundary. One also notes that the transformation
(\ref{symm*}) brings the 
Fourier
image $S_k$ into its complex conjugate $S_k^*$,  so that the invariance of 
the function 
$S_i$ under the symmetry (\ref{symm*}) implies that $S_k \in \mathbb{R}$.

To discretize the problem~\eqref{appB:eq},  
one introduces the uniform lattice
with sites at the points $\alpha_n = \Delta(n + 1/2)$, 
where $\Delta = \pi/N$,  $n = 0,\dots, N-1$. 
The boundary condition~\eqref{bc} 
takes the form
\begin{subequations}
\label{system}
\begin{equation}
\label{bc1}
(\phi_f)_{n=0} = 0\;.
\end{equation}
The condition at the other boundary of the orbifold, 
$\alpha=\pi$, is obtained 
in the following way. 
One discretizes Eq.~\eqref{eqphi} in a straightforward manner 
and writes down a finite--difference
equation relating $(N-1)$-th and $N$-th sites of the circle. 
Then, the values of the 
fields at the $N$-th site are expressed in terms of the values at 
$(N-1)$-th
site by using the orbifold--symmetry 
condition, and
one obtains
\begin{multline}
\label{bc2}
\frac{1}{\Delta}\big[\pi - 2(\phi_f)_{N-1} + 2\Re (S_i)_{N-1}\big] = \\
1+\frac{\mu T}{2\sin^2(\alpha_{N-1/2})}
\sin\big(\Re(S_i)_{N-1} + (\phi_f)_{N-1} - \alpha_{N-1}\big)
\cosh\big(\Im(S_i)_{N-1}\big)\;.
\end{multline}
When discretizing Eq.~\eqref{eqphi} in the internal points 
of the orbifold, one
has to deal with the explicit singularity at $\alpha = 0$
in the right hand side of this equation.
After several attempts, we have
found a discretization scheme which does not 
produce numerical artifacts at $\alpha=0$.
It has the following form
\begin{multline}
\label{eqn}
\frac{1}{\Delta}\big[(S_i)_n - (S_i)_{n-1}\big] - 
\frac{1}{\Delta}\big[(\phi_f)_n - (\phi_f)_{n-1}\big] = \\1
+ \frac{\mu T}{2 \sin^2(\alpha_{n-1/2}/2)}
\left\{\sin\big( (S_i)_{n-1} + (\phi_f)_n - \alpha_{n-1/2}\big)
+ \frac{\Delta}{2}\cos\big( (S_i)_{n-1} + (\phi_f)_n - 
\alpha_{n-1/2}\big)\right\}\\
- \frac{\mu^2 T^2 \Delta^2}{4\sin^4(\alpha_{n-1/2}/2)}
\sin\big((S_i)_n + (\phi_f)_n - \alpha_{n-1/2}\big)
\cos\big((S_i)_{n-1} + (\phi_f)_n - \alpha_{n-1/2}\big)\;.
\end{multline}
\end{subequations}
Here the index $n$ runs from $1$ to $N-1$. It is 
straightfoward to check that 
Eq.~\eqref{eqn} for $\alpha_n\sim 1$ provides a 
second--order approximation of equation~\eqref{eqphi}. 
At the same time, for $\alpha_n \sim \Delta$ the last term of 
Eq.~\eqref{eqn} becomes dominant, and one obtains
\begin{equation}
\nonumber
(S_i)_n + (\phi_f)_n - \alpha_n = 0 \;\;\; \mathrm{when} \;\;\; \
\alpha_n \sim \Delta\;.
\end{equation}
Again, the discrepancy between this equation and the 
continuous one is $O(\Delta^2)$.
The approximation turns out to be worse in the region 
between $\alpha_n\sim\Delta$ and 
$\alpha_n\sim 1$. Careful numerical analysis of the scaling
of error with   
the lattice spacing shows that the approximation order 
lies somewhere in between 1 and 2.

Finally, the Fourier transform~\eqref{fourSi} is 
replaced by its discrete version, with 
the sum taken from $k=0$ to $k=N-1$. We choose the real
variables $S_k$ and $(\phi_f)_n$ as the unknowns. 
Since $k,n = 0\dots(N-1)$, the 
total number of unknowns, $2N$, coincides with the number of real
equations in the system (\ref{system}) (two real boundary conditions
(\ref{bc1}), (\ref{bc2}) and $(N-1)$ complex equations
(\ref{eqn})). We solve the system of equations (\ref{system}) by 
the Newton--Raphson method. 
Then, the values of the final energy
and imaginary part of the action are given by the formulae 
obtained by a straighforward discretization
of Eqs. (\ref{actionreal}), (\ref{Efinal}). 
In order to calculate numerically the initial 
energy (\ref{Einitial}), we note that 
$$
\tilde{R_i'}(z=iT) = \frac{1}{2T}\big[1-iS_i'(w=0)\big] = 
\frac{1}{2T}\big[1+S_{k=1}\big]\;.
$$
The comparison between 
the values of the initial and final energies yields the  
estimate of the discretization error.
In our calculations we used the lattice with $2048$ sites, 
and the discretization error was smaller than $10^{-3}$ for all our
solutions.

\section{Solutions at small $T$}
\label{app:C}

The purpose of this Appendix is twofold. First, we are going to 
clarify the structure of the solution of the $T/\theta$-problem
at small values of the parameter $T$.
Second, we will obtain
the formulae for the suppression exponent 
in the region of energies $2.3E_S<E<E_l$ which is not covered by
the numerical method of Sec.~\ref{sec:jumps1}.
To make analysis more transparent, we work below 
in the limit $\mathrm{e}^{-\theta}=0$ and set $\mu=1$.

First we observe that the limiting solution found in
Sec.~\ref{sec:limiting} represents the first terms of a systematic
expansion of the solution in powers of the parameter $T$.
We start with the ``hard'' core.
One notices that the characteristic length scale in the ``hard''
region is set by the parameter $T$.
So, it is natural to represent the solution in this region in the
following form,
\be
\label{hardexp}
\phi_{i,f}^{(h)}=\phi_{i,f}^{(h0)}(\zeta)+T\phi_{i,f}^{(h1)}(\zeta)
+T^2\phi_{i,f}^{(h2)}(\zeta)+\ldots\;,
\ee
where $\zeta$ is a rescaled variable,
$$
    \zeta = z/T.
$$
We allow the functions
$\phi_{i,f}^{(h0)}(\zeta),\phi_{i,f}^{(h1)}(\zeta),\ldots$ to
contain terms of order $(\ln{T})^n$. To illustrate this point
let us write the ``hard'' core 
solution of Sec.~\ref{sec:limiting} in terms of the
rescaled variable~$\zeta$. 
Making use of Eqs.~\eqref{varphisol}, \eqref{rsol}, \eqref{Ri}, 
\eqref{matching}, one obtains,
\begin{subequations}
\label{appC:phi}
\begin{align}
\phi_i^{(h)} =&  \left\{ \frac{\pi}{2} + i\ln\left(\frac{T}{2}\right) + 
i\ln(\zeta-i)\right\}\notag\\& +
  T\left\{2 i A \ln{T} + 2 i A \frac{\zeta+i}{\zeta-i}\ln
\left(\frac{\zeta+i}{2 i}\right) - A\zeta - i(A+B)\right\}
\label{appC:phii}\\
\phi_f^{(h)} =& \frac{\pi}{2} +
 T \left\{
-A(\zeta - i)\ln(\zeta - i) - A (\zeta + i)\ln(\zeta+ i) 
- 2 A\zeta \ln\left(\frac{T}{2}\right) + B\zeta  - \pi A \label{appC:phif}
\right\}\;,
\end{align}
\end{subequations}
where the coefficients $A$ and $B$ are defined in
Eqs.(\ref{nearzero}). 
It is clear that 
these expressions represent the first two terms of the 
expansion (\ref{hardexp}). 

In the ``soft'' region the characteristic length scale is of order
1. So, the expansion in the ``soft'' region has the form 
\be
\label{softexp}
\phi_{i,f}^{(s)} = \phi_{i,f}^{(s0)}(z) + T \phi_{i,f}^{(s1)}(z) + 
T^2\phi_{i,f}^{(s2)}(z)+\dots\;.
\ee
The first term of this expansion has been found in
Sec.~\ref{sec:limiting}. 

To construct the solution in the entire complex plane one should match the
asymptotics of the expansion (\ref{hardexp}) at $\zeta\to\infty$ with
the Taylor series of the expansion (\ref{softexp}) at $z\to 0$. Note that
because the arguments in the functional expansions (\ref{hardexp}) and
(\ref{softexp}) differ by rescaling, matching intertwines terms of
different order in $T$ in the expansions (\ref{hardexp}),
(\ref{softexp}). Thus, in Sec.~\ref{sec:limiting} we have used
the behavior of the zeroth order approximation in the ``soft'' region to
determine the coefficients of the ``hard'' part of the solution to the
first order in $T$, see Eqs.~(\ref{matching}). Here we proceed to
calculate the first order term in the ``soft'' part of the solution.

Since $\mathrm{e}^{-\theta}=0$, the function
$\phi^{(s1)}_{i}$ is regular in the upper half--plane of the 
``soft'' region. Thus, its real and imaginary parts on the real axis, 
$a^{(1)}(x)$ and $b^{(1)}(x)$,
are related by the Cauchy formula~\eqref{aprime}. 
Equations~\eqref{bprime},~\eqref{uprime} in the first order 
approximation are linearized:
\label{appC:eq}
\begin{align}
\label{appC:eq1}
&\left[\frac{d}{dx} - c(x)\right] b^{(1)}  =  -s(x) u^{(1)},\\
\label{appC:eq2}
&\left[\frac{d}{dx} + c(x)\right] u^{(1)}  =  2 
\frac{da^{(1)}}{dx} - s(x) b^{(1)}\;.
\end{align}
Here, as usual, $u^{(1)} = a^{(1)} + \phi_{f}^{(1)}$, while the coefficients
$c(x)$, $s(x)$ are 
\begin{eqnarray}
\nonumber
c(x) = \ch b^{(0)} \cos u^{(0)}\;,\;\;\; s(x) = \sh b^{(0)} \sin u^{(0)}\;.
\end{eqnarray}
Boundary conditions at $x=0$ originate from matching 
with the ``hard'' part of the solution. One writes 
expressions~\eqref{appC:phi}
in terms of the variable $z = T\zeta$ and extracts the 
corrections proportional
to $T$, by formally considering the variable $z$ as a quantity of order one.
According to the above discussion, this fixes
the small--$z$ behavior of the functions $\phi^{(s1)}_{i,f}$:
\[
\begin{array}{l}
\phi_i^{(s1)} \to \left\{\frac{1}{z} + 2 iA
\ln\left(\frac{z}{2}+i0\right) 
- i(A+B)  +  \pi A \right\}\\ 
\phi_f^{(s1)} \to -\pi A
\end{array}\;\;\;\;\mathrm{at}\;\;\; z\to 0\;,
\]
Finally, we recall that due to Eqs.~\eqref{boundreal*} 
the finctions $a^{(1)}$, $b^{(1)}$, $u^{(1)}$ vanish at $x\to +\infty$. 

The problem outlined above can be solved by iterative method 
similar to that of Sec.~\ref{sec:limiting}. 
Namely, one starts from $u^{(1)}=1/x$ and 
finds the functions $b^{(1)}$, $a^{(1)}$, $u^{(1)}$ 
by solving Eqs.~\eqref{appC:eq1},~\eqref{aprime},~\eqref{appC:eq2}
one after another. In this way the improved approximation for $u^{(1)}$ is
obtained,  
and the next cycle of iterations begins. After 30 iterations the solution
is found numerically with accuracy 
of order $10^{-4}$.

We have determined the solution in the first--order approximation both in 
the ``soft'' and ``hard'' regions. So, we are able to calculate its energy
by performing the integration in Eq.~\eqref{Einitial0} in the 
first--order approximation. We find
\begin{equation}
\label{appC:E}
E(T) = E_l + 4\pi A T\left(2 A\ln{T}-B\right) + W T\;.
\end{equation}
In this expression $E_l$ stands for the limiting value,  
the second term is the contribution of the first--order corrections 
in the ``hard'' core, and the last term represents the correction calculated 
numerically in the ``soft'' region,
\[
W = 4\int\limits_0^{+\infty}dx\; \left[\frac{d\phi_f^{(0)}}{dx} 
\frac{d\phi_f^{(1)}}{dx}\right]=9.6\;.
\]
Given the expression (\ref{appC:E}) for the energy, it is straightforward
to 
evaluate the suppression exponent
using Eqs.~\eqref{ENLegendre}, \eqref{F},
\begin{equation}
\label{appC:F}
F\big|_{N=0} = F_l - T^2 [f_1\ln T + f_2]\;,
\end{equation}
where $F_l$ is the suppression at the limiting energy; the 
constants $f_1$, $f_2$ are
related to the coefficients of Eq.~\eqref{appC:E}. Numerically, 
$f_1 = 17.9$, $f_2 = 49.5$. 

\begin{sloppy}
We use the expressions~\eqref{appC:E}, \eqref{appC:F} to determine the
function  
$F(E)$ in the region ${2.3E_S<E<E_l}$. Note, though, that
the expressions~\eqref{appC:E}, \eqref{appC:F} are valid only in the vicinity
of the point ${E=E_l}$; corrections to these formulae grow when one goes
away from the limiting energy. 
At $E=2.3 E_S$ we are able to compare the value of 
the suppression exponent given by
Eqs.~\eqref{appC:E},~\eqref{appC:F} with the one calculated numerically 
by the method of Sec.~\ref{sec:jumps1}. 
We have found that these two values coincide
with the precision of order 0.5\%.

\end{sloppy}

\end{document}